# Computing Apparent Planetary Magnitudes for *The Astronomical Almanac*

Anthony Mallama  (corresponding author)

14012 Lancaster Lane, Bowie, MD, 20715, USA

anthony.mallama@gmail.com

James L. Hilton

US Naval Observatory, 3450 Massachusetts Avenue NW, Washington, DC, 20392, USA

james.l.hilton1@navy.mil

Pages:         64

Tables:        10

Figures:       14

Revised: 2018 June 21

Abstract

Improved equations for computing planetary magnitudes are reported. These formulas model *V*-band observations acquired from the time of the earliest filter photometry in the 1950s up to the present era. The new equations incorporate several terms that have not previously been used for generating physical ephemerides. These include the rotation and revolution angles of Mars, the sub-solar and sub-Earth latitudes of Uranus, and the secular time dependence of Neptune. Formulas for use in *The Astronomical Almanac* cover the planetary phase angles visible from Earth. Supplementary equations cover those phase angles beyond the geocentric limits. Geocentric magnitudes were computed over a span of at least 50 years and the results were statistically analyzed. The mean, variation and extreme magnitudes for each planet are reported. Other bands besides *V* on the Johnson-Cousins and Sloan photometric systems are briefly discussed. The planetary magnitude data products available from the U.S. Naval Observatory are also listed. An appendix describes source code and test data sets that are available on-line for computing planetary magnitudes according to the equations and circumstances given in this paper. The files are posted as supplementary material for this paper. They are also available at SourceForge under project https://sourceforge.net/projects/planetary-magnitudes/ under the 'Files' tab in the folder 'Ap_Mag_Current_Version'.

Keywords: Planets and satellite: general, ephemerides

1. Introduction

Apparent magnitudes are an essential element of planetary physical ephemerides. They are generally computed along with other physical quantities such as the sub-solar latitude and the phase angle. An associated measure called 'surface brightness' (usually given in magnitudes per square arc-second) is critical for planning observations where an exposure time must be computed in advance. One example is remote observation from spacecraft where commands must be uploaded ahead of the planned observation. Brightness data are required any time that a signal-to-noise ratio is needed. Apparent magnitudes are also widely listed in almanacs such as *The Astronomical Almanac*, magazines intended for amateur astronomers, newspaper articles for the general public, as well as in astronomical observers' guides and astronomy textbooks. More recently, on-line ephemerides such as the [U.S. Naval Observatory's Topocentric Configuration of Major Solar System Bodies](USNO), [HORIZONS](Giorgini et al. 1996) and self-contained software such as the Multiyear Interactive Computer Almanac (MICA 2005) have also been providing apparent planetary magnitudes. Finally, the direct detection of exo-planets depends on their apparent magnitudes, which may be estimated from their solar system counterparts.

In most cases the apparent magnitude refers to that on the 'visual system'. The *visual magnitude* is an old term referring to observations made with the human eye principally during the era before electronic sensors became available. Nowadays the visual magnitude is commonly taken to mean the *V*-band of the Johnson-Cousins photometric system ([Johnson and Morgan, 1953](); [Cousins, 1976a]() and [Cousins 1976b]()). The response curve of that band is centered at 0.549 µm and has a full-width-at-half-maximum of 0.086 µm. Thus, it is somewhat like the response curve of the human eye. Magnitudes herein are taken to be on the *V*-band unless otherwise indicated.

The purpose of this paper is to specify formulas for computing apparent magnitudes of solar system planets based upon the latest models and the most complete sets of observations available. Section 2 briefly reviews the literature on planetary magnitudes and then describes how apparent magnitudes are computed. Section 3 discusses the apparent magnitude of each planet individually and presents the equations for computing their apparent brightness. Section 4 lists statistics of the apparent magnitudes, such as mean opposition values, brightest and faintest and the greatest brilliancy of Venus. Section 5 provides an overview of other wavelength bands besides *V* in the Johnson-Cousins system that may be useful to observers. The Sloan photometric system is also described because it is becoming the new standard. Section 6 lists the planetary magnitude data products that are available from the U.S. Naval Observatory. Section 7 summarizes the paper and presents our conclusions. An appendix describes source code and test data sets for computing planetary magnitudes. The files are hosted on-line at SourceForge which is an open-source software site.

2. Planetary magnitudes

Müller (1893) developed Eq. 1, a general-purpose formula for predicting apparent magnitudes of the planets. The apparent magnitude depends upon the planet's distance from the Sun, *r*, and from the Earth, *d*, in accordance with the inverse square law. Another important factor is the illumination phase angle, *α*, which is defined as the arc between the Sun and the sensor with its vertex at the planetocenter. Thus, small values of *α* correspond with more fully illuminated disks and large values of *α* to thin crescents.

$$V = 5 \log_{10}( r\, d ) + V_1(0) + C_1 \alpha + C_2 \alpha^2 + \ldots$$

Eq. 1

*V* is the apparent visual magnitude, and $V_1(0)$ is the magnitude when observed at *α* = 0 and when the planet is at a distance of one au from both the Sun and the observer. $V_1(0)$ is sometimes referred to as the planet's *absolute magnitude* or *geometric magnitude* and it may also be thought of as $C_0 \alpha^0$. The sum $\Sigma_n C_n \alpha^n$ is called the phase function. The phase function generally increases the planet's apparent magnitude with increasing phase angle.

Nearly the entire 180° of the phase curves for Mercury and Venus have been observed. Fig. 1 shows that the brightness of airless Mercury declines dramatically with phase angle while that of cloud-covered Venus drops off less sharply. Mars and the Earth are intermediate cases between the extremes of Mercury and Venus. The Earth-viewable ranges of *α* for the giant planets are restricted, from about 12° for Jupiter to less than 2° for Neptune. The phase functions of Jupiter and Saturn have been determined accurately over their entire observable ranges. The magnitude changes for Uranus and Neptune as seen from the Earth arising from phase angle is less than 0.01 magnitude. So, the phase functions can be ignored for the purposes of computing apparent geocentric ephemeris magnitudes for these two planets. The phase functions of the giant planets for large values of *α*, shown in Fig. 2, are based upon measurements obtained from interplanetary spacecraft.

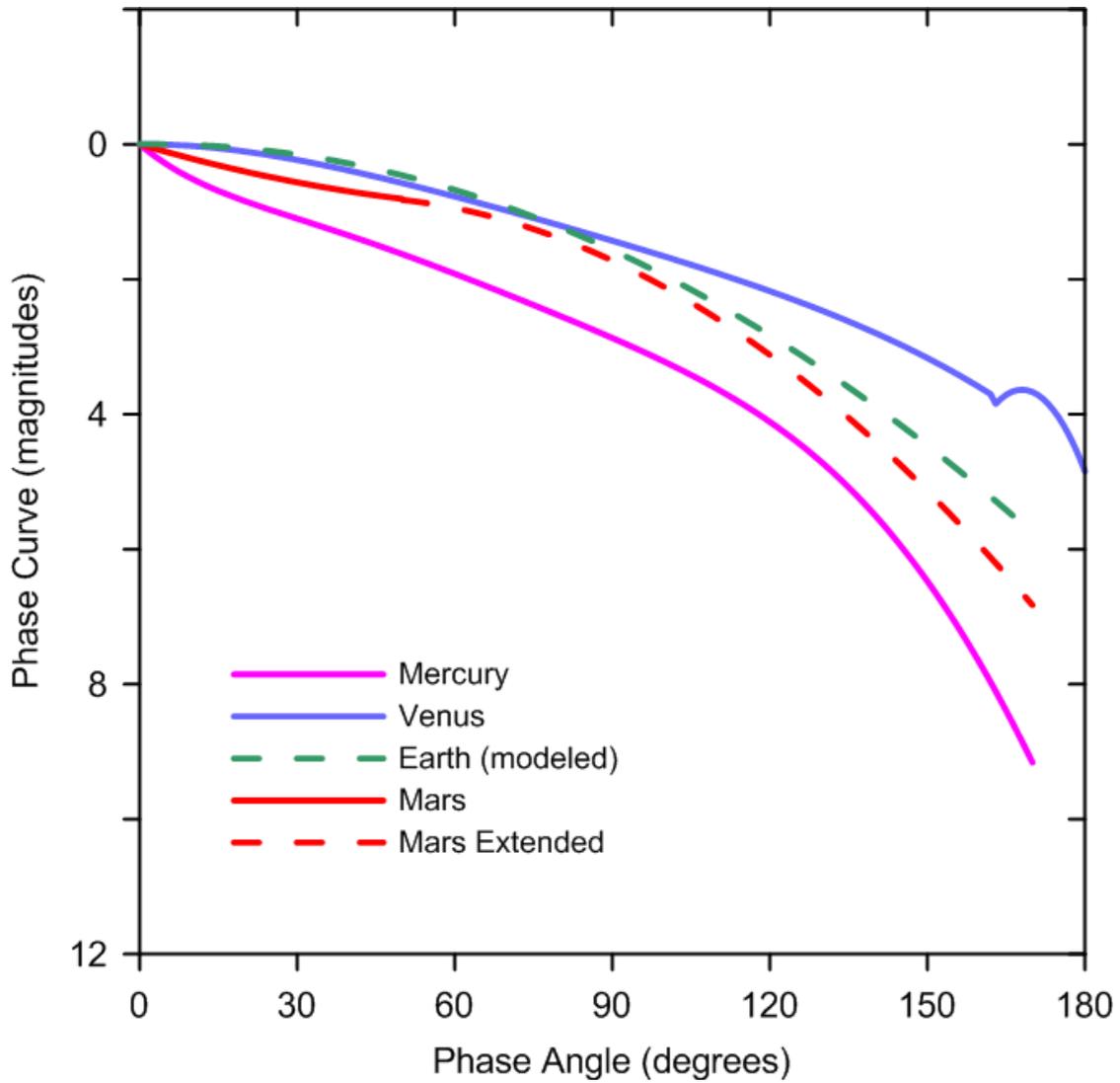

*Fig. 1. The phase curve for Mars with its thin atmosphere lies between those of barren Mercury and cloud-shrouded Venus. The upturns in the phase functions of Mercury and Mars near α = 0° are due to strong backscattering from their surfaces. Forward scattering by liquid droplets in Venus' atmosphere is the source of the inflection point in the phase curve of Venus at about 163°. The dashed lines for the Earth and Mars are discussed in section 3.3 and 3.4, respectively.*

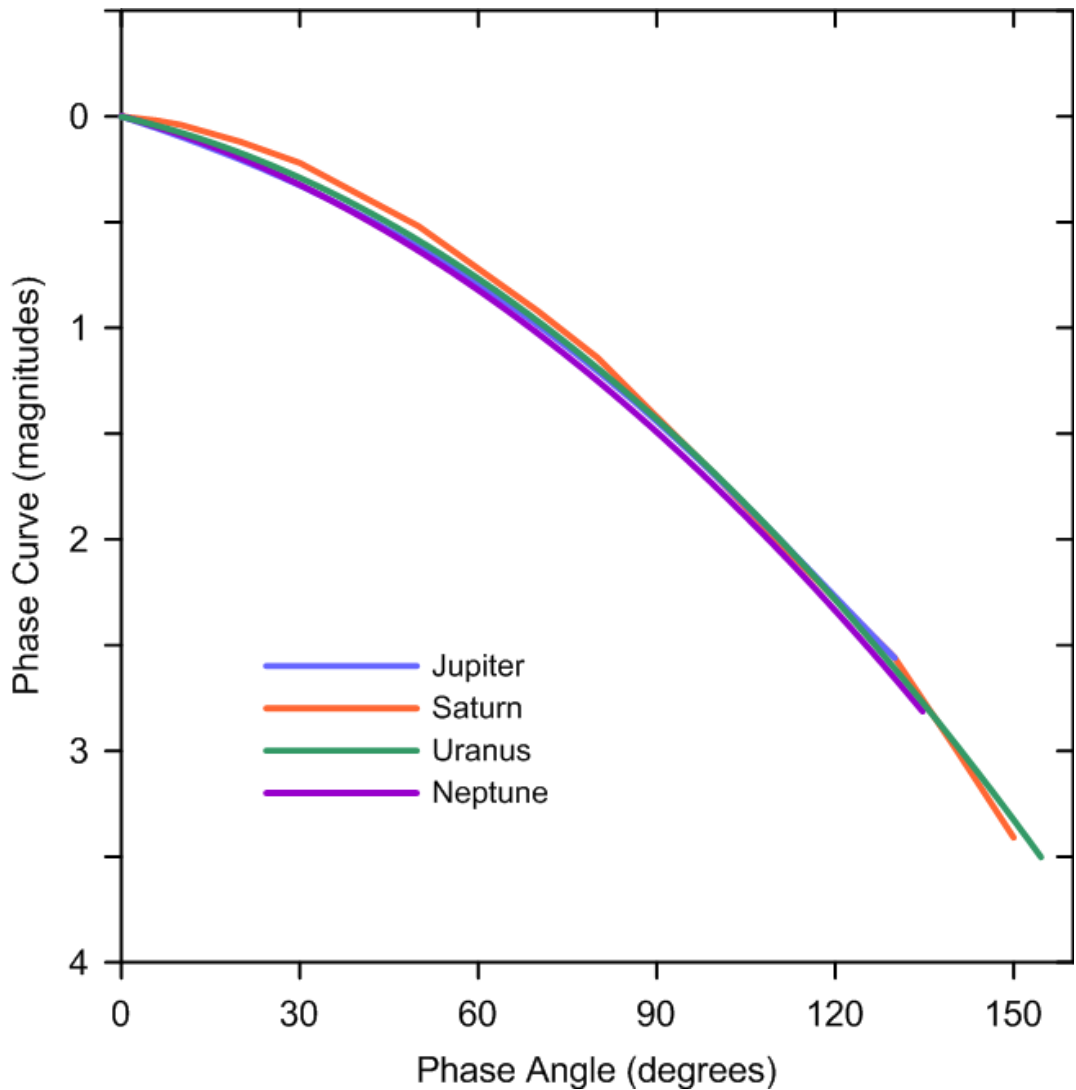

*Fig. 2. The phase curves for the giant planets are more similar to one another than are those of the terrestrial bodies. The large phase angle data are based upon observations obtained from interplanetary spacecraft as explained in section 3.5 through 3.8.*

More than 50 years ago Harris (1961) summarized the available observations and analyses of planetary brightness. Those studies began in the 1800s with visual magnitudes actually estimated by the human eye. Filtered photoelectric photometry only became standardized when Johnson and Morgan (1953) specified the characteristics of the *V*-band and established a set of reference stars with accurately calibrated magnitudes. Only the earliest *V*-band magnitudes of the planets were available to Harris.

The equations given by Harris are now outdated. They do not capture all the significant aspects of planetary brightness and variability observed in the six decades since his work was published. Subsequent observations include a wider range of physical and geometrical conditions as well as the longer time span of observation. Many subtleties also remained hidden until more accurate observations were obtained from space-based instruments and from ground-based telescopes using CCD sensors.

Mallama et al. (2017) reported up-to-date values of $V_1(0)$ for all the planets and listed the best fitting coefficients of $\alpha$. They also evaluated more observational circumstances and listed coefficients for other parameters such as rotation angle, inclination and time period, where needed. The following synopsis mentions a few of the more striking geophysical implications especially as they pertain to apparent magnitudes.

Mercury was observed with the SOlar and Heliospheric Observatory (SOHO) spacecraft, which allowed coverage of a greater range of phase angles than ever before, extending over $2^o < \alpha < 170^o$. Mallama et al. (2002) analyzed those magnitudes along with ground-based CCD data. Mercury's large brightness surge near $\alpha = 0^o$ was attributed to coherent backscattering from its regolith. Furthermore, the complex shape of the planet's phase curve indicates a surface that is about as rough as that of the Moon.

SOHO also recorded magnitudes of Venus. These observations along with ground-based CCD data covered a range in phase angle of $2^o < \alpha < 179^o$. An anomalous brightness excess in the phase curve near $\alpha = 163^o$ was found to be due to sunlight forward scattered by sulfuric acid droplets above the main cloud deck (Mallama et al. 2006). There is also a brightness reversal as $\alpha$ approaches $0^o$, which is due to a glory arising from the planet's atmosphere (Muñoz et al., 2014).

Mallama (2007) demonstrated that Mars' apparent magnitude is a function of the viewing geometry that includes the sub-Earth point, the sub-solar point, and the albedo markings on the Martian surface. There is also a seasonal dependence, that is a variation arising from the angle between Mars' orbital longitude and the node of its vernal equinox on its orbit.

Changing intensities of Jupiter's cloud belts correlate with variations of its integrated magnitude (Mallama and Schmude, 2012). However, these changes are too small to be relevant to ephemeris predictions.

Saturn's ring system contributes greatly to the brightness variations of the whole system. Thus, its magnitude depends on their inclination angle ([Schmude 2011](#) and [Mallama 2012](#)).

The brightness of Uranus, like Saturn, varies with its inclination. However, the changes are due to the depletion of light-absorbing methane towards its poles and correspondingly greater reflection of sunlight when the planet is more inclined to the Earth and Sun ([Schmude et al. 2015](#)).

Finally, the brightness of Neptune has increased markedly since electronic monitoring began over 60 years ago. The geophysical reason for this change is still unknown ([Schmude et al. 2016](#)), but recent observations provide a better basis for computing the planet's ephemeris brightness at the present time than do the older data.

3. Individual Planets

The following sub-sections address each of the eight planets separately. The relevant parameters, such as *r*, *d* and *α* are indicated and formulas for *V* are provided. All angular coefficients use units of degrees.

3.1 Mercury's apparent magnitude

Hilton (2005) and Mallama et al. (2002, referred to herein as 'the Mercury paper') analyzed the photometry of Mercury using third- and seventh-order polynomials, respectively. We have re-analyzed the observations for this paper and find that a sixth-order polynomial provides a better fit to the data than does the third-order and that it fits as well as the seventh-order. Therefore, we have adopted the new sixth-order polynomial in Eq. 2.

$$V = 5 \log_{10} (r\,d) - 0.613 + 6.3280\text{E}{-}02\ \alpha - 1.6336\text{E}{-}03\ \alpha^2$$
$$+ 3.3644\text{E}{-}05\ \alpha^3 - 3.4265\text{E}{-}07\ \alpha^4 + 1.6893\text{E}{-}09\ \alpha^5 - 3.0334\text{E}{-}12\ \alpha^6$$

Eq. 2

There are two issues pertaining to the *V1(0)* term, -0.613, that are worth noting. One is that this magnitude is about 8% fainter than the result derived from geophysical modeling of the photometry as reported in the Mercury paper. This brighter $0^\text{o}$ phase angle magnitude results from fitting the opposition surge with the physical model described there. The other issue is that the zeroth order coefficient in Table A-1.2 of Mallama et al. (2017) is incorrect. That value (-0.694) comes from the geophysical solution and it should not have been combined with the polynomial solution for orders 1 through 7.

3.2 Venus' apparent magnitude

Hilton (2005) used the observations acquired and later published by Mallama et al. (2006, referred to herein as 'the Venus paper') to improve the magnitudes in *The Astronomical Almanac*. The Venus paper fit these data with a polynomial of degree four. Hilton used a polynomial of degree three which did not provide enough flexibility to capture the brightness reversal at small values of $\alpha$ where Venus actually becomes *dimmer* as $\alpha$ approaches 0. This unusual phenomenon is captured by the fourth-order polynomial in the Venus paper. Muñoz et al. (2014) were the first to recognize the significance of this brightness reversal and identified it physically as a glory (Laven, 2005) caused by Venus' atmosphere. Another brightness reversal occurs at $\alpha = 163°$. This phenomenon was identified in the Venus paper as forward scattering of sunlight by droplets of $H_2SO_4$ high in the atmosphere of Venus. The inflection in the phase curve is abrupt and discontinuous as shown in Fig. 1. Thus, a two-part piecewise function is required to parameterize the phase curve. Eq. 3 and Eq. 4 are from the Venus paper and are valid in the intervals of $0° < \alpha \leq 163.7°$ and $163.7° < \alpha < 179°$, respectively.

$$V = 5 \log_{10}(r\,d) - 4.384 - 1.044\text{E-}03\,\alpha + 3.687\text{E-}04\,\alpha^2 - 2.814\text{E-}06\,\alpha^3 + 8.938\text{E-}09\,\alpha^4$$

Eq. 3

$$V = 5 \log_{10}(r\,d) + 236.05828 - 2.81914\text{E+}00\,\alpha + 8.39034\text{E-}03\,\alpha^2$$

Eq. 4

3.3 Earth's apparent magnitude

Values of *V* in this paper and in *The Astronomical Almanac* refer to magnitudes of other planets for an observer located at the geocenter. In the case of the Earth itself though, one must assume that the observer is located well above the Earth's surface.

Mallama et al. (2017) established a value of $V_1(0)$ = -3.99 based upon an analysis of spectrophotometry from the EPOXI spacecraft as reported by Livengood et al. (2011). Meanwhile, Tinetti et al. (2006) modeled the albedo of the Earth as a function of $\alpha$ for a variety of cloud conditions and terrain types. Mallama et al. fit the albedos of the 'realistic clouds' case in figure 7 of Tinetti et al. with a spline function. Eq. 5 is the resulting polynomial phase curve in magnitudes (represented by the green dashed line in Fig. 1) of the spline combined with the value of $V_1(0)$.

$$V = 5 \log_{10} ( r\ d ) - 3.99 - 1.060\text{E-}3\ \alpha + 2.054\text{E-}4\ \alpha^2$$

Eq. 5

Tinetti et al.'s geometric ($\alpha = 0^o$) albedos range by a factor of more than 6 from 0.12 in the cloud-free case to 0.76 in the case of alto-stratus clouds, which indicates that the Earth can be quite variable in its intrinsic brightness. The $V_1(0)$ magnitude from Mallama et al. corresponds to an albedo of 0.434 while that of Tinetti et al. for the 'realistic cloud' case and wavelengths from 0.5 to 0.9 µm is 0.358. These values bracket the albedo derived from the $V_1(0)$ magnitude quoted in Tholen et al. (2000). The Earth's apparent magnitude probably cannot be predicted as accurately as those for most of the other planets given these uncertainties.

3.4 Mars' apparent magnitude

The brightness of Mars depends on the sub-Earth longitude of its illuminated disk according to Mallama (2007, referred to herein as 'the Mars paper'). The planet is brightest when the highly reflective Amazonis-Tharsis region ($\lambda \sim 115^{o}$) is near the sub-Earth longitude, and it is faintest when the large, low-reflectance feature Syrtis Major ($\lambda \sim 290^{o}$) is near the sub-Earth longitude. The root-mean-square variation over longitude in the *V*-band is about 0.035 magnitude with excursions as large as 0.060 magnitude. Therefore, an accurate estimate for the apparent magnitude of Mars requires the sub-Earth and sub-solar longitude which, taken together, give an effective sub-longitude, $\lambda_e$, of the visible sunlit hemisphere. The angle $\lambda_e$ changes rapidly, approximately $351^{o}$ day$^{-1}$, so correct determination of the effect arising from $\lambda_e$ requires an accurate knowledge of the observer's time. The time-dependent error can be as large as 0.04 mag hr$^{-1}$. The dependence on $\lambda_e$ may not be included in hardcopy tabulations such as *The Astronomical Almanac* for this reason.

The Mars paper tabulates the contribution from the sub-longitude at ten-degree intervals. Interpolation at intermediate effective longitudes is accomplished here using a polynomial representation of Stirling's interpolation formula using differences up to fourth-order (Duncombe 2013, §14.2.4). This method was chosen so the predicted integrated apparent magnitude of the disk will be a smooth, continuous function with the current predicted value being a function of the preceding and following longitudinal sectors of the planet along with the current sub-longitude. The effect of this correction for the sub-Earth longitude was compared to 64 high accuracy *V* band observations of Mars by Young (1974). Including the correction reduced the RMS scatter of the observations by 0.019 magnitude.

The Mars paper also indicates that Mars' brightness depends on the longitude of its vernal equinox on its orbit, $L_s$. This change in brightness is partially caused by the change in viewing angle arising from the sub-Earth latitude which can be as great as $30.3^{o}$ from Mars' equator. Furthermore, Geissler (2005) demonstrates that there are seasonal changes primarily arising from changes in Mars' polar caps. Geissler also documents aperiodic changes arising from concealing and revelation of features from dust transported by storms. Thus, the data are noisy. Finally, there are gaps in the coverage in all photometric bands. $L_s$ is a slowly changing angle, about $0.52^{o}$ day$^{-1}$, so its effect on the predicted apparent magnitude is more easily tabulated in *The Astronomical Almanac*. The same algorithm used for interpolating the effect of $\lambda_e$ is used to estimate the contribution of $L_S$ to the predicted magnitude. The predicted magnitude in the gap in the data at $L_S = 240°$ is estimated by choosing the value that minimized the difference between the fourth order interpolation and a linear interpolation. The justification for this choice is that the contribution that would have been observed is not independent of the contribution from those of nearby, observed values of $L_S$. This choice represents the best that can be gleaned from those nearby values. The estimated uncertainty in the true contribution near $L_S = 240°$ is only slightly larger than at other values of $L_S$. There was no detectable change in the RMS scatter of the test observations when the $L_S$ correction is included either with or without the correction for the longitude.

Finally, the mean decrease from global dust storms on Mars in the *V*-band is −0.12 magnitude. Such storms are rare and are difficult to predict, so this phenomenon is not considered here.

The formula for computing *V* according to the Mars paper is given here as Eq. 6 and it is valid for $\alpha \leq 50°$, which includes the full range of phase angles visible from Earth.

$$V = 5 \log_{10} ( r\ d ) - 1.601 + 0.02267\ \alpha - 0.0001302\ \alpha^2 + L(\lambda_e) + L(L_S)$$

Eq. 6

where *L($\lambda_e$)* and *L($L_S$)* are the magnitude corrections for the longitude of the sub-Earth meridian of the illuminated disk and the longitude of the vernal equinox, respectively.

The phase curve of Mars with its thin atmosphere beyond $\alpha = 50°$ may be approximated by averaging the dimming in magnitudes for airless Mercury (Eq. 2) and that for the Earth (Eq. 5). The resulting polynomial is combined with the *L($\lambda_e$) and L($L_S$)* functions to give Eq. 7

$$V = 5 \log_{10} ( r\ d ) - 0.367 - 0.02573\ \alpha + 0.0003445\ \alpha^2 + L(\lambda_e) + L(L_S)$$

Eq. 7

where the constant value '-0.367', is chosen to match the value of Eq. 6 at $\alpha = 50°$. This formula (represented by the red dashed line in Fig. 1) for the phase curve at large values $\alpha$ of should give reasonable estimates for the apparent magnitudes of Mars. However, they are not expected to be as accurate as those for Mercury and Venus, especially beyond $\alpha \sim 120^{\circ}$ where the phase functions for Mercury and the Earth begin to strongly diverge.

3.5 Jupiter's apparent magnitude

The phase curve of Jupiter as seen from Earth cannot exceed $\alpha = 12^o$, so its relatively uncomplicated phase function can be reproduced by a second-order polynomial. Mallama and Schmude (2012, referred to herein as 'the Jupiter paper') analyzed all the available *V* magnitude observations of this planet. They detected changes of a few hundredths of a magnitude in Jupiter's intrinsic brightness, which appear to be related to variations in the cloud bands. However, such changes are difficult to predict, so they are not modeled in the magnitude equation. Eq. 8, taken from the Jupiter paper, applies to $\alpha \leq 12^o$.

```
V = 5 log₁₀ ( r d )  - 9.395 – 3.7E-04 α + 6.16E-04 α²
```
Eq. 8

Mayorga et al. (2016) determined the phase curve of Jupiter beyond $\alpha = 12^o$ based on observations from the ISS instrument on the Cassini spacecraft. They modeled the functions for several filters with fifth-order polynomials. The polynomial coefficients of the planet's albedo in the green filter listed in their table 2 are used in Eq. 9 to represent the phase curve at larger values of $\alpha$. Mayorga et al. state that the function is not trustworthy beyond α~130$^o$ because there are no data to constrain that region. Eq. 8 explicitly covers $0 \leq \alpha < 12^o$ so, Eq. 9 should be taken to represent *V* in the range $12^o < \alpha < 130^o$. The value -9.428 in Eq. 9 is an adjustment to -9.395 from Eq. 8 so the two equations agree at $\alpha = 12^o$. The complete phase curve is shown in Fig. 2 of this paper.

```
V = 5 log₁₀ ( r d ) – 9.428 –2.5 log₁₀ (1.0 – 1.507 * (α/180.) – 0.363 * (α
/180.)**2 –0.062 * (α /180.)**3 + 2.809 * (α/180.)**4 –1.876 * (α/180.)**5)
```
Eq. 9

Note that the logarithm function is required to convert from the albedo measures used by Mayorga et al. to magnitudes and that the phase angles are divided by 180$^o$ in accordance with their parameterization.

3.6 Saturn's apparent magnitude

The apparent brightness of Saturn together with its ring system depends strongly on the inclination of the ring plane to both the observer and the Sun. Saturn's rings increase the brightness of the overall planetary system and complicate the empirical determination of *V* for the globe alone because they must be measured together in most circumstances. The phase function for the rings is different from that of the globe, as shown by Fig. 3, and thus adds complexity to the modeling of *V* for the entire system. Finally, the planet and the ring system can also both occult and eclipse portions of one another.

The *planetocentric* latitudes, $β_E$ and $β_S$, represent the inclination of the rings as seen from the Earth and from the Sun, respectively. Mallama (2012, referred to herein as 'the Saturn paper') used these latitude values as indicated in Eq. 10 which applies to $α < 6.5°$ and $β < 27°$ for the planet and rings.

$$V = 5 \log_{10}(r\,d) - 8.914 - 1.825 \sin β + 0.026\, α - 0.378 \sin β \, e^{(-2.25\, α)}$$

Eq. 10

The effective inclination, *β*, is $(β_E β_S)^{1/2}$ when $β_E$ and $β_S$ have the same sign, and *β*= 0, when $β_E$ and $β_S$ have contrary signs. The latter case covers a rare condition where the Sun lights one side of the rings and the observer sees the other side, so the rings are backlit and very faint.

*The Astronomical Almanac* and MICA tabulate the magnitudes of Saturn including its rings system, but the HORIZONS System currently reports magnitudes of the globe without the rings. As stated earlier in this section, the brightness of Saturn's globe alone is difficult to measure because of the presence of the rings in most photometric measurements. Mallama and Pavlov (2017) addressed this problem by deriving synthetic magnitudes from the spectrophotometry of Karkoschka (1998), which he obtained during the ring-plane-crossing of 1995 when the rings were practically invisible. (A synthetic magnitude is the integral of the product of spectral flux and instrumental response over the frequency range of the filter band-pass.)  The $V_1(0)$ magnitude for Saturn was determined to be -8.95. That study also showed that setting *β* = 0° in Eq. 10 has a first-order phase curve slope of 0.026 magnitude / degree. This slope is unrealistically steep for the globe alone, and is probably due to aliasing from the much steeper slope of the ring system. Therefore, this paper uses an alternative method to model *V* for Saturn's globe at $α < 6.5°$. It adopts the second-order polynomial for Jupiter and combines that with the synthetic magnitude for Saturn. The resulting formula is given in Eq. 11. While Eq. 10 would indicate a dimming of 0.17 magnitude at α < 6.5$^o$ Eq. 11 indicates a much more reasonable 0.02 magnitude for the globe alone.

```
V = 5 log₁₀ ( r d ) - 8.95 - 3.7E-04 α + 6.16E-04 α²
```

Eq. 11

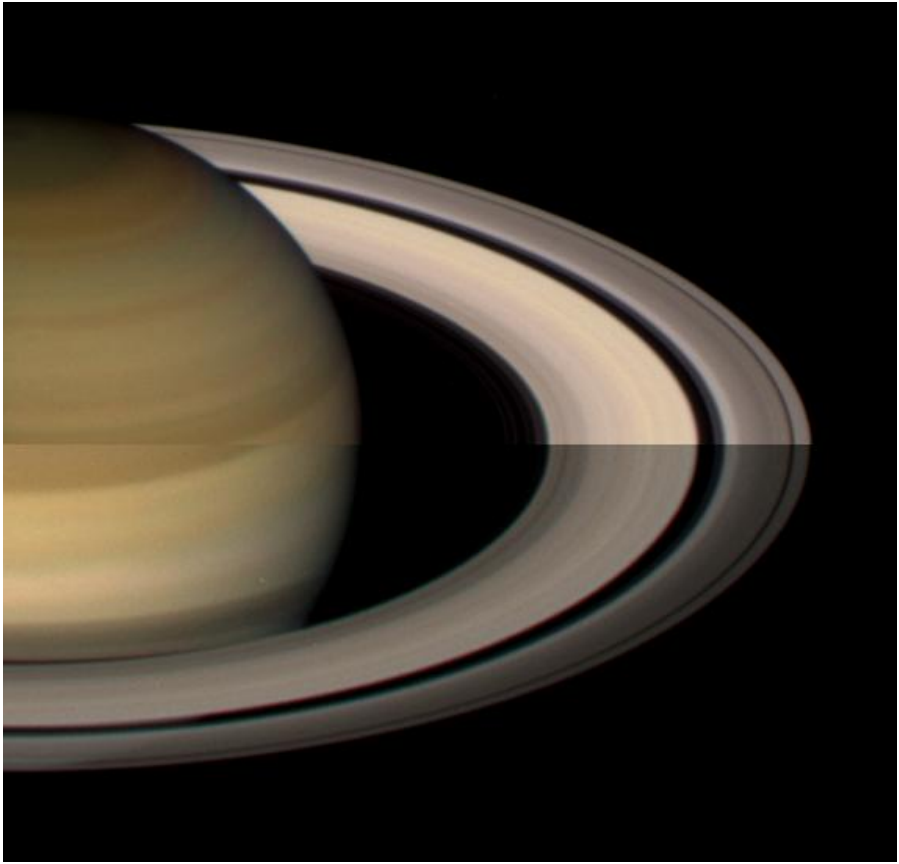

*Fig. 3. The brightness of Saturn's rings near α = 0$^o$ (top half of image) is greatly enhanced as compared to that at α = 6$^o$ (bottom half). Meanwhile the brightness of the globe is nearly the same at both phase angles. Hubble Space Telescope images acquired in the F439W (blue), F555W (green) and F675W (red) filters were used to generate this composite. The figure was originally published in the Saturn paper.*

Finally, Dyudina et al. (2005) modeled the albedo of Saturn over a wide range of $\alpha$ based upon anisotropic scattering functions derived from Pioneer spacecraft data by earlier analysts. Their figure 7, which illustrates the phase function for red light, was digitized and albedo was converted to magnitude. (There are no filter data corresponding to the green *V*-band but their figure 4 indicates that the scattering phase function for blue light is similar to that for red. So, green is probably similar as well.) Then a fourth-order polynomial was fit to the magnitudes as indicated in Eq. 12. Fig. 2 illustrates that the dimming from this function is similar to that of the other giant planets. The zeroth order coefficient of −8.94 provides agreement with Eq. 11 at $\alpha = 6^{\circ}$.

$$V = 5 \log_{10}(r\,d) - 8.94 + 2.446\text{E-}4\,\alpha + 2.672\text{E-}4\,\alpha^2 - 1.505\text{E-}6\,\alpha^3 + 4.767\text{E-}9\,\alpha^4$$

Eq. 12

Eq. 12 may be used to compute an approximate *V* magnitude for the globe of Saturn only when $6^{\circ} < \alpha < 150^{\circ}$. There is not enough information available at present from which to derive an equation for the apparent magnitude of Saturn including its rings for $\alpha > 6.5^{\circ}$.

3.7 Uranus' apparent magnitude

The pole of the rotational axis of Uranus is inclined 82° to its orbit and rotates in the retrograde direction, so the sub-solar latitude ranges from +82° to –82°. Schmude et al. (2015, referred to herein as 'the Uranus paper') determined that the planetographic sub-Earth, $\varphi'_E$, and sub-solar latitudes, $\varphi'_S$, have a substantial effect on the apparent magnitude. The conversion from planetocentric latitude, $\varphi$, to planetographic latitude is

$$\varphi' = \tan^{-1}\frac{\tan\varphi}{(1-f)^2}$$

Eq. 13

where $f$ is the flattening of the planet. For Uranus $f$ = 0.0022927. Uranus reflects light more strongly when the polar regions are near the center of its visible disk because those latitudes are depleted in light-absorbing methane. The planetographic latitude is the complement of the angle between the body's spin axis and a line perpendicular to the ellipsoid at a given point on its surface, while the planetocentric latitude is the angle subtended at the center of the ellipsoid between the spin axis direction and the direction to a point on the surface.

Because Uranus' distance from the Sun is much larger than the Earth's, the maximum value of $\alpha$ is only 3.2°. The effect of such a small phase angle on magnitude is negligible. Lockwood (1978) reported a change of only 0.003 magnitude due to phase angle. Likewise, evaluation of the well-established phase curve of Jupiter (which has similar light-scattering properties) at 3.2° (Eq. 7) gives a change of 0.005 magnitude. Therefore, $\alpha$ is not a significant factor in computing apparent $V$ for Uranus as seen from the Earth. Eq. 14 for computing the planet's apparent brightness is from the Uranus paper.

```
V = 5 log₁₀ ( r d ) - 7.110 - 8.4E-04 φ'
```

Eq. 14

where $\varphi'$ is the average of the absolute values of $\varphi'_E$ and $\varphi'_S$.

Pearl et al. (1990) constructed the phase curve of Uranus over the large range of $\alpha$ illustrated in their figure 3 using measurements from the radiometer (0.30 – 1.78 µm) on the Voyager 2 spacecraft in addition to other Voyager data published by Pollack et al. (1986). The pairs of $\alpha$ and data number (DN) values in that figure were digitized and the DN values were converted to magnitudes of dimming. Finally, a second-order polynomial was fit to the magnitudes and normalized to zero at $\alpha$ = 0°. Fig. 2 of this paper shows that the dimming for Uranus is similar to that of the other giant planets. Eq. 15 combines $V_1(0)$ and the dependence on $\varphi'$ with the second order polynomial. The data plotted by Pearl et al. extends to 154° so the equation should be valid to that limit. When evaluated at the maximum phase angle of Uranus that is visible from Earth, $\alpha$ = 3.1°, the dimming is 0.021 magnitude. This is somewhat larger than expected and is probably due to the lack of observational constraints for $\alpha$ < 15.8° in figure 3 of Pearl et al.

```
V = 5 log₁₀ ( r d ) - 7.110 - 8.4E-04 φ' + 6.587E-3 α + 1.045E-4 α²
```
Eq. 15

3.8 Neptune's apparent magnitude

Neptune was observed in the *V* band from 1954 through 1966 and from 1991 through 2014 as reported by Schmude et al. (2016, referred to herein as 'the Neptune paper'). Furthermore, Lockwood (http://www2.lowell.edu/users/wes/U_N_lcurves.pdf) and Karkoschka (2011) list medium-band *y* magnitudes reduced to the planet's mean opposition distance, which fill in the gap from 1966 through 1991. The effective wavelength of the *y* band is similar to that of *V* and adjustment of the two magnitudes axes shown in Fig. 4 results in good correspondence between the two bands. This figure illustrates that the planet brightened significantly between about 1980 and 2000. Since that time though, the brightness of Neptune has been relatively constant. The albedo of Neptune is now close to that of Uranus while it was much lower prior to 1980.

The Neptune paper lists two slopes for the change of the *V* magnitude over time. The first, −0.00223 per year, is for the entire 1954-2014 span of observations. The second, -0.00377 per year, is for a more recent and more intensive period of observations covering 1993-2014. However, neither of these represents the relatively constant brightness before 1980, the increase from 1980 through 2000 or the brightness plateau after 2000 illustrated by the combination of *V* and *y* magnitudes in Fig. 4. Eq. 16 is based on the new analysis in this paper of both *V* and *y* magnitudes which models Neptunian magnitudes separately for the pre-1980, 1980-2000, and post 2000 time periods

$$V = \begin{cases} 5 \log_{10}(r\,d) - 6.89 & t < 1980.0 \\ 5 \log_{10}(r\,d) - 6.89 - 0.0054\,(t - 1980) & 1980.0 \leq t \leq 2000.0 \\ 5 \log_{10}(r\,d) - 7.00 & t > 2000.0 \end{cases}$$

Eq. 16

where *t* is the CE (Common Era) year.

While the future brightness of Neptune may be computed with Eq. 16, the unknown cause of the planet's variability casts doubt upon its long-term predictive accuracy. Two hypotheses have been put forth to explain the intrinsic brightness changes. Karkoschka (2011) suggested that Neptune experiences occasional 'darkening events' where haze particles are elevated to higher levels in the atmosphere. Brightening as the dark aerosols settle to lower levels follows these 'darkening events'. On the other hand, Sromovsky et al. (2003) suggested that the variability is a seasonal effect because the brightening occurred during the spring season in the planet's southern hemisphere, which is similar to what occurs on Uranus. Lockwood and Jerzykiewicz (2006) criticized that interpretation noting that earlier observations were not well represented by the seasonal model. Fig. 4 plots the observed brightness and the sub-solar latitude of Neptune since the 1950s when photometric monitoring in standardized band-passes began. We point out that Neptune reached its southern solstice in 2005 and that the next equinox will occur in 2046. The recent time period during which the apparent magnitude has been constant is approximately bisected by the southern summer solstice. So, the planet should soon begin to fade if the seasonal model is correct.

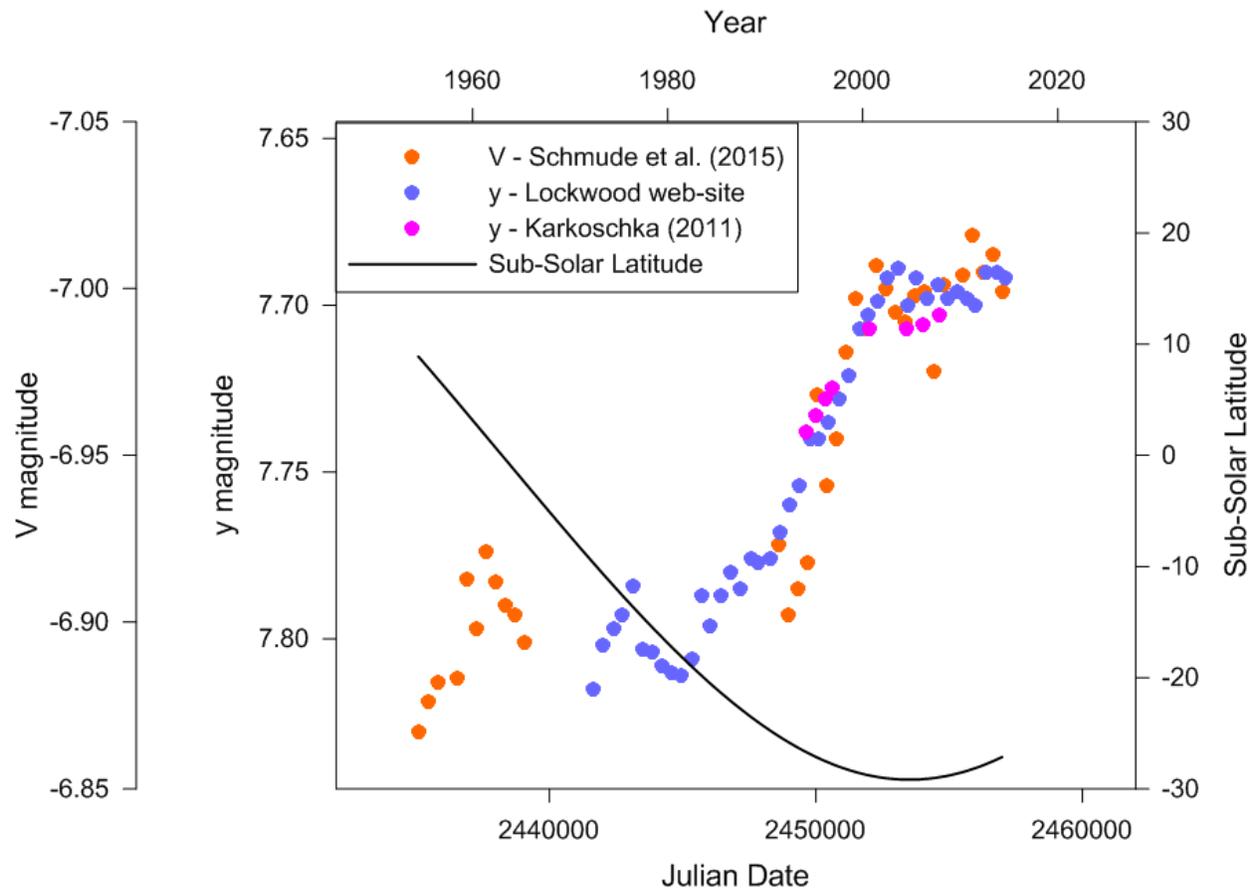

*Fig. 4. Wide-band, V, and medium-band, y (yellow), magnitudes plotted along with the sub-solar latitude. Neptune increased in brightness during its southern spring season. The y-band data of Lockwood from his web-site* http://www2.lowell.edu/users/wes/U_N_lcurves.pdf *is used with permission.*

The very large distance of Neptune from the Sun limits the maximum value of $\alpha$ as seen from Earth to 1.9°. Karkoschka (2011) measured a phase angle coefficient of just 0.0028 magnitudes per degree. The value inferred from the phase curve of Jupiter at 1.9° is less than 0.01 magnitude. In either case, the effect of the phase angle on the planet's apparent magnitude is only a few milli-magnitudes and, thus, it is ignored.

Pearl and Conrath (1991) constructed the phase curve of Neptune over a wide range of $\alpha$ illustrated in their figure 4 using measurements from the same Voyager 2 radiometer employed with Uranus and also including earlier data published by Pollack et al. (1986). The method of digitizing and processing their data for the analysis in the present paper was the same as that for Uranus. The data extend to $\alpha$ = 134.4°, however the polynomial in this paper fits a slightly smaller limit of $\alpha$ = 133.14° because that is the last 'direct full-disk measurement'. Fig. 2 demonstrates that the resulting phase curve for Neptune is similar to those of Jupiter, Saturn and Uranus. Eq. 17 combines $V_1(0)$ with the second order polynomial for Neptune shown in Fig. 2 and it is valid to $\alpha$ = 133°. When evaluated at the maximum phase angle of Neptune visible from Earth, $\alpha$ = 1.9°, the dimming is 0.015 magnitude. This is larger than expected and may be due to the lack of observational constraints for $\alpha$ < 13.9° in figure 4 of Pearl et al. The constant, –7.00, is from Eq. 16 and it indicates that this equation pertains to $t$ > 2000.0.

$$V = 5 \log_{10}(r\,d) - 7.00 + 7.944\text{E-}3\ \alpha + 9.617\text{E-}5\ \alpha^2$$

Eq. 17

## 4. Statistics of the apparent V magnitudes

The following sub-sections report the brightest, faintest, mean and other statistical values related to *V* for each planet as seen from the Earth. Unique geometrical circumstances relating to the extremes and variations of brightness are discussed. We also break down the magnitude variations according to their causes. These include the Sun-planet and Earth-planet distances for every planet and the phase functions for most of them. Other causes such as the angles of inclination, rotation, and vernal equinox longitude are included for the planets to which they apply. The final sub-section discusses the extremes of the magnitude and their variations over all the planets.

In order to thoroughly assess the statistics of apparent magnitudes we have computed their values at daily intervals over periods of at least 50 years. These time spans are centered on 2017.0 and the magnitudes are computed at 0 hours UT unless otherwise noted. Distances and geometries are based upon accurate orbital positions for the planets, the Earth and the Sun. Precise physical ephemeris data related to specific planets, such as longitude of the sub-Earth meridian of the illuminated disk of Mars, are also used as needed. When planets are occulted by the Sun their magnitudes are not computed. Magnitudes corresponding to transits of the Sun are excluded for Mercury because the planet would be too faint to be predicted accurately. Venus remains relatively bright during solar transits, so its magnitude is computed and discussed as a special case.

Time scales for predictable changes in the apparent magnitudes of the planets range from several minutes for Mars' rotation in sub-Earth longitude to decades for Neptune. The exact time intervals chosen for analysis depend on the *synodic* period, that is, the mean time required for the apparent geocentric ecliptic longitude of the planet to repeat. During a synodic period the phase angles and the Earth-planet distances approximately repeat. The use of an integer number of synodic periods assures that computed mean magnitudes and standard deviations are not skewed by uneven sampling of the geometry. For more distant planets the time intervals are based more strongly on their periods of revolution around the Sun. During that time the Earth may complete many circuits itself and, thus the geometry also repeats. Analyzing one or more complete revolution of the distant planets additionally takes into account their perihelion and aphelion distances. It also covers the range of sub-latitude angles, which contribute to the planet's apparent brightness. For example, the model for Saturn requires latitude information to compute the inclination of its rings system.

The phase function for very small and very large values of $\alpha$ for Mercury and Venus are not observable due to these planets' small angular separation from the Sun near superior and inferior conjunctions.

(For values of $\alpha > 179.3^o$ Mercury is transiting the Sun and for $\alpha < 0.7^o$ it is occulted by the Sun. The equivalent values for Venus are $\alpha > 179.6^o$ and $\alpha < 0.4^o$.) Therefore, extrapolation was required to determine some of their magnitudes. We present results that include all daily magnitudes (some of which required extrapolation) and separate results for just those magnitudes corresponding to values of $\alpha$ within the observed range.

The *variations* reported below correspond to the individual components of the equations for the apparent magnitudes. For example, that for the Earth-planet distance contributes to the apparent magnitude according to the inverse-square law. These components are computed individually and then combined to form the apparent magnitude.

Finally, each sub-section includes a plot of the relative brightness of the planet as a function of its elongation from the Sun. This gives an indication of what time of day or night may be most advantageous for viewing.

4.1 Mercury's magnitude statistics

There are 157.60 synodic periods of Mercury in 50 years. The planet's highly eccentric orbit (e = 0.206) highlights the importance of sampling an integer number of synodic periods. So, the 50-year time span was increased by 47 days to make an even 158 synodic periods.

Figure 5 displays the apparent magnitude of Mercury over this period as a function of apparent elongation from the Sun. Analysis of the resulting computed apparent magnitudes from Eq. 2 indicates that Mercury is brightest when observed near $\alpha = 0^o$. The planet is then at superior conjunction and, thus, is farthest from the Earth. This somewhat surprising outcome, that the influence of small phase angle dominates the inverse square law, is due to the brightness surge known as the 'opposition effect'. Shkuratov et al. (1999) attribute this phenomenon to coherent backscatter from the planet's regolith, while its shadow hiding aspect is addressed by Lynch and Livingston (1995). The overall effect was modeled in the Mercury paper.

The brightest of the 151 superior conjunction magnitudes over the 50-year period of analysis was −2.48 on 2006-May-19 when $\alpha = 1.17^o$ and Mercury was near perihelion at r = 0.310 AU. The mean of the superior conjunction magnitudes was -1.89. These and other statistics of the planet's apparent magnitude are listed in the first part of Table 1.

Mercury is faintest when it is backlit by the Sun near inferior conjunction. The values of $\alpha$ at inferior conjunction ranged from about 169 to 179$^o$ over the period of analysis and the mean apparent magnitude is +5.93. The faintest magnitude was found to be +7.25 on 2029-May-13 when the $\alpha = 179.13^o$.

The mean *V* magnitude of Mercury is +0.23 and its standard deviation of 1.78 is the greatest of any planet. The large standard deviation is due mostly to the strong effect of $\alpha$ on brightness as discussed later in this section.

The brightest and faintest magnitudes listed in the first part of Table 1 depend on extrapolation beyond the observed range of phase angles, which was $2.1^o < \alpha < 169.5^o$. The next part of the Table lists the brightest and faintest magnitudes for values of $\alpha$ that are within that observed range. The standard deviation is smaller for data without extrapolation. The mean magnitude is brighter when no extrapolated data are used because very faint magnitudes around the time of inferior conjunction are omitted.

Table 1. Statistics of the *V* magnitude for Mercury

```
                  Apparent magnitudes

  With extrapolation (0.7 – 179.2 deg)
Brightest                             -2.48
Faintest                              +7.25
Mean                                  +0.23
Standard Deviation                     1.78
Standard Deviation of the Mean         0.01
Mean Superior Conjunction             -1.89
Mean Inferior Conjunction             +5.93

 Without extrapolation (2.1 - 169.5 deg)
Brightest                             -2.43
Faintest                              +5.64
Mean                                  +0.12
Standard Deviation                     1.60
Standard Deviation of the Mean         0.01

                  Variations by component

            With extrapolation
Earth-Mercury distance                 2.11
Sun-Mercury distance                   0.91
Both distances                         2.56
Phase angle                           10.82

            Without extrapolation
Earth-Mercury distance                 2.10
Sun-Mercury distance                   0.91
Both distances                         2.55
Phase angle                            9.08

Magnitudes evaluated                 18,303
Start date                       1991-Dec-08
Stop date                        2042-Jan-23
```

The variations corresponding with individual components of the equation for the apparent magnitude include the Earth-Mercury distance, the Sun-Mercury distance and the phase angle. The greatest variation is from phase angle (10.82 magnitude with extrapolation and 9.08 without) and it far exceeds those arising from the Earth-Mercury distance (2.11 with or 2.10 without) and the Sun-Mercury distance (both 0.91). The variation due to the combined solar and terrestrial distances is 2.56 with extrapolation and 2.55 without. These variational results are listed in the third and fourth parts of the Table.

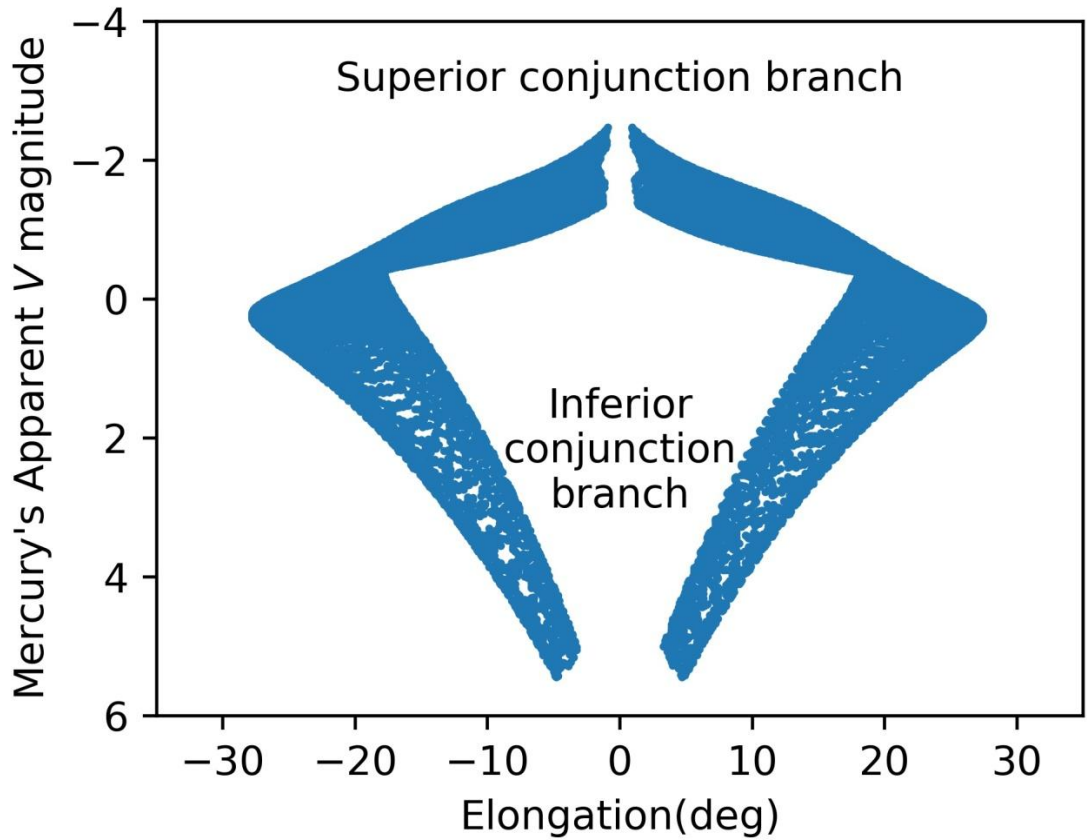

*Fig. 5. The apparent magnitude of Mercury is spread out over approximately eight magnitudes, but only 27.7$^o$ in elongation from the Sun. The brighter (upper branch) of the distribution occurs when Mercury is near superior conjunction. Mercury is on the far side of the Sun from the Earth, so the size of its apparent disk is small, but nearly fully lighted. There is a definite 'opposition effect' near superior conjunction.*

4.2 Venus' magnitude statistics

The synodic period of Venus is 583.92 days. Five of those periods equals 7.997 Earth years. During that interval Venus revolves around the Sun 13 times while the Earth revolves nearly 8 times and, therefore, the Earth-Venus-Sun geometry almost repeats itself. This nearly commensurate period is ideal for even sampling of magnitudes. Therefore 7 periods totaling 55.96 years were evaluated. Eq. 3 was used to compute $V$ when $\alpha \leq 163.7^{\circ}$ and Eq. 4 was used when $\alpha > 163.7^{\circ}$ for the reasons given in Section 3.2.

The phase curve of Venus is less steep than that of Mercury, as shown in Fig. 1, while the distance from the Earth to Venus varies more than that from the Earth to Mercury. Therefore, distance affects the apparent magnitude of Venus more strongly than does the phase angle and, in fact, Venus is *not* brightest near superior conjunction where $\alpha = 0^{\circ}$. Analysis of the daily magnitudes indicate that Venus reaches its greatest brilliancy, on average, at $\alpha = 123.50^{\circ} \pm 1.31^{\circ}$. (The visible hemisphere of Venus is only 22% illuminated at that value of $\alpha$.) The mean apparent magnitude at greatest brilliancy is -4.81 ± 0.07 and the mean elongation of the planet from the Sun is $37.08^{\circ} \pm 0.59^{\circ}$. Venus was brightest over the 50 year period of analysis on 1989-Dec-19 at magnitude -4.92 and when $\alpha = 124.15^{\circ}$ and the distance between Venus and the Earth, $d$, was only 0.377 AU. At its peak, the brightness of Venus exceeds that of any other planet as seen from the Earth by nearly two magnitudes. Its mean magnitude, -4.14, is also the brightest of all the planets. The variation of magnitude with elongation from the Sun is illustrated in Fig. 6.

Venus remains very bright near the time of its inferior conjunction because forward scattering in its atmosphere re-directs sunlight toward the observer as discussed in the Venus paper. Special computations were performed for the date of 2004-Jan-08/09 when Venus transited the Sun. At mid-transit, when $\alpha = 179.7^{\circ}$, the estimated magnitude was -2.98 which is taken to be its faintest value given the current geometry of the Earth's and Venus's orbits.

The standard deviation of the magnitudes, 0.31, is much less than that of Mercury because of the reduced effect of phase angle on the brightness of Venus. The statistics of the apparent magnitude of Venus are listed in the first part of Table 2.

The extreme values for Mercury depended fairly strongly on whether magnitudes computed for dates when the planet was outside of the range of observed phase angles are included. For Venus, on the other hand, nearly the whole phase curve (2.0 to $179.0^{\circ}$) has been observed. Thus, all but one of the statis-tics that include extrapolated magnitudes (listed in the top portion of Table) applies to the case of 'no extrapolation'. The one exception is the faintest magnitude where the value -3.14 corresponding to $\alpha = 178.92^o$ on 1996-Jun-11 is listed in the second part of the Table.

The variations of the apparent magnitude for Venus due to distance are nearly equal to those due to phase angle. The third and fourth parts of the Table indicate 4.09 magnitudes of variation for distance, while that for phase angle is 4.69 in the extrapolation case and 4.64 without extrapolation. This contrasts with Mercury where the effect of phase angle strongly outweighs that from distance. Venus' orbit is significantly less eccentric than Mercury's, so the change in brightness from Venus-Sun distance is much less.

Table 2. Statistics of the *V* magnitude for Venus

```
                  Apparent magnitudes

  With extrapolation (1.4 – 179.7 deg)
 Brightest                              -4.92
 Faintest                               -2.98
 Mean                                   -4.14
 Standard Deviation                      0.31
 Standard Deviation of the Mean          0.01
 Mean Greatest Brilliancy               -4.81

 Without extrapolation (2.0 – 179.0 deg) *
  Faintest                              -3.14

                 Variations by component

             With extrapolation
 Earth-Venus distance                    4.08
 Sun-Venus distance                      0.03
 Both distances                          4.09
 Phase angle                             4.69

             Without extrapolation *
 Phase angle                             4.64

 Magnitudes evaluated                  20,423
 Start date                        1989-Jan-10
 Stop date                         2044-Dec-22
```

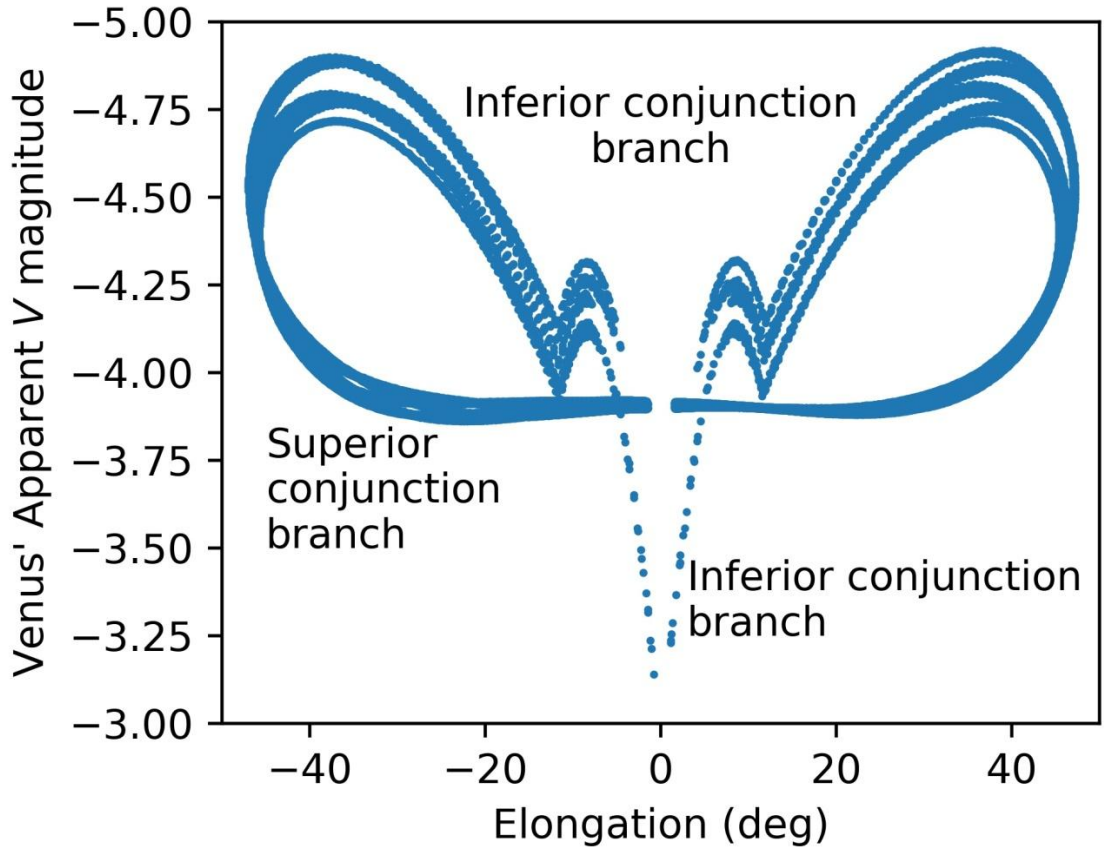

*Figure 6. The predicted apparent magnitude of Venus is generally smaller when it is nearer inferior conjunction (between the Earth and the Sun) than when it is nearer superior conjunction. The forward scattering of light causes the sudden decrease in magnitude at an elongation about 11.5$^o$ from inferior conjunction by sulfuric acid in Venus' atmosphere. The slight asymmetry between the eastern (positive) and western elongations is likely an artifact of the near 13:8 commensurability between the Earth's orbital period and Venus' synodic period with the Earth and will probably dissipate as the period of time covered increases.*

4.3 Earth's magnitude statistics

Values of *V* in Section 4 refer to magnitudes of other planets as seen from the Earth. Therefore, statistics of the Earth's apparent magnitude in that sense are undefined. To get some sense of the Earth's magnitude, though, one may assume that the observer is located elsewhere in space.

The *V* magnitude for the Earth was calculated during the 50-year period centered on 2017.0. When the Earth is at opposition to the Sun as seen from Venus on 2038-Jan-04 the predicted value of *V* is astonishingly bright at -6.91. This brightness is primarily due to the small distance of 0.265 AU between Venus and the Earth. The faintest, -2.76, was on 1992-Jun-14 when the Earth was in conjunction with the Sun and, thus, on the opposite side from Venus at a distance of 1.736 AU.

Thus, the minimum magnitude of the Earth seen from Venus is nearly 2 magnitudes less than the minimum magnitude of Venus seen from the Earth, while the maximum magnitude of the Earth seen from Venus is only about 0.25 magnitudes less than the maximum magnitude of Venus seen from the Earth. This difference in the range arises, in part, because the full Earth is lighted at its maximum brilliance while only 22% of Venus' surface is illuminated.

From Mars, the Earth was brightest at *V* = -2.55 on 2005 July 30 when $\alpha$ = 95.89$^o$. The faintest magnitude cannot be accurately determined because the albedo function given by [Tinetti et al. (2006)](#) appears to approach zero as $\alpha$ approaches 180$^o$. During a lunar eclipse the Moon's surface appears a red or copper color from light forward scattered through the Earth's atmosphere. So, the Earth's albedo, like Venus, remains positive at large phase angles, but there is a significant change in its integrated apparent color.

The Earth would be progressively more difficult to observe from the giant planets. For innermost Jupiter and outermost Neptune its maximum elongation from the Sun would be only about 11.1$^o$ and 1.9$^o$, respectively. The corresponding apparent *V* magnitudes would be 1.1 and 5.0.

4.4 Mars' magnitude statistics

The mean synodic period of Mars is 779.96 days but the considerable eccentricity (e = 0.093) of its orbit makes the Earth-Mars distances vary significantly from period to period. Furthermore, the ratio between the synodic period of Mars and its own orbital period results in a second period of 15.78 Earth years. In practice, this means that Mars is nearest to our planet approximately every 15 or 17 years, that is, either 7 or 8 synodic periods.

Analysis of 100 years of daily distances centered on 2017.0 indicated that the 62 years beginning on 1988-Sep-23 (one day after a closest approach) and ending on 2050-Aug-16 (a day of closest approach) represent an evenly distributed sample of Earth-Mars distances for statistical evaluation. Fig. 7 shows the distribution in apparent magnitude as a function of apparent elongation over this period. Besides the usual variables $r$, $d$ and $\alpha$, the apparent magnitude of Mars depends on the values of rotational longitude, $\lambda$, and longitude of the vernal equinox, $L_S$. Fig. 7a displays the apparent magnitudes without the effect of $\lambda$ and $L_S$, while Fig. 7b includes them. The effects of $\lambda$ and $L_S$ were used to determine extrema in the predicted apparent visual magnitude for Mars as described by Eq. 6.

The brightest magnitude, -2.94, is predicted to occur on 2050-Aug-15, which is one day before an approach of just 0.374 AU. Mars is faintest when it is on the far side of the Sun from the Earth but *not* when it is exactly in conjunction with the Sun. The planet exhibits a moderate opposition surge so it is slightly *brighter* near solar conjunction than it is before or after when the Earth-Mars distance is smaller. The faintest magnitude during this period is $V$ = 1.86 on 2036-Jul-09 when $\alpha$ = 15.06$^o$. The mean value of $V$, 0.71, is much closer to its least brightness than to its greatest brightness because Mars only spends a small amount of time near the Earth during each synodic period. The standard deviation of $V$, 1.05, is smaller than that of Mercury but larger than any other planet. These and other statistical values are reported in the first part of Table 3.

The apparent magnitude at mean opposition for an outer planet can be estimated by taking $r$ to be the semi-major axis, $a$, of the planet's orbit, and taking $d$ to be $a - 1$ because the semi-major axis of the Earth's orbit is one au. In the case of Mars, the mean opposition magnitude calculated by this method is -2.09. This is the 'nominal' value listed in the Table. When the daily magnitudes are analyzed, the mean at opposition is -1.98 with a standard deviation of 0.57. The mean is slightly fainter than the nominal value because Mars spends more time near aphelion than near perihelion. The standard deviation is rather large due to the moderately eccentric Martian orbit.

The variations of Martian magnitudes by component are shown in the second part of the Table. The largest of these is the combined Sun-Mars and Earth-Mars distance effect, which can produce a variation of 4.69 magnitudes. The variation due to phase angle is only 0.78 magnitude because only a small range of angles is visible from the Earth. The variations from orbital and rotational effects are 0.15 and 0.11, respectively.

Table 3. Statistics of the *V* magnitude for Mars

```
                  Apparent magnitudes

Brightest                                  -2.94
Faintest                                   +1.86
Mean                                       +0.71
Standard Deviation                          1.05
Standard Deviation of the Mean              0.01
Mean Opposition (Nominal)                  -2.09
Mean Opposition (Statistical)              -1.98

                  Variations by component

Earth-Mars distance                         4.28
Sun-Mars  distance                          0.41
Both distances                              4.69
Phase angle                                 0.78
Orbital                                     0.15
Rotational                                  0.11

Magnitudes evaluated                      22,600
Start date                            1988-Sep-23
Stop date                             2050-Aug-16
```

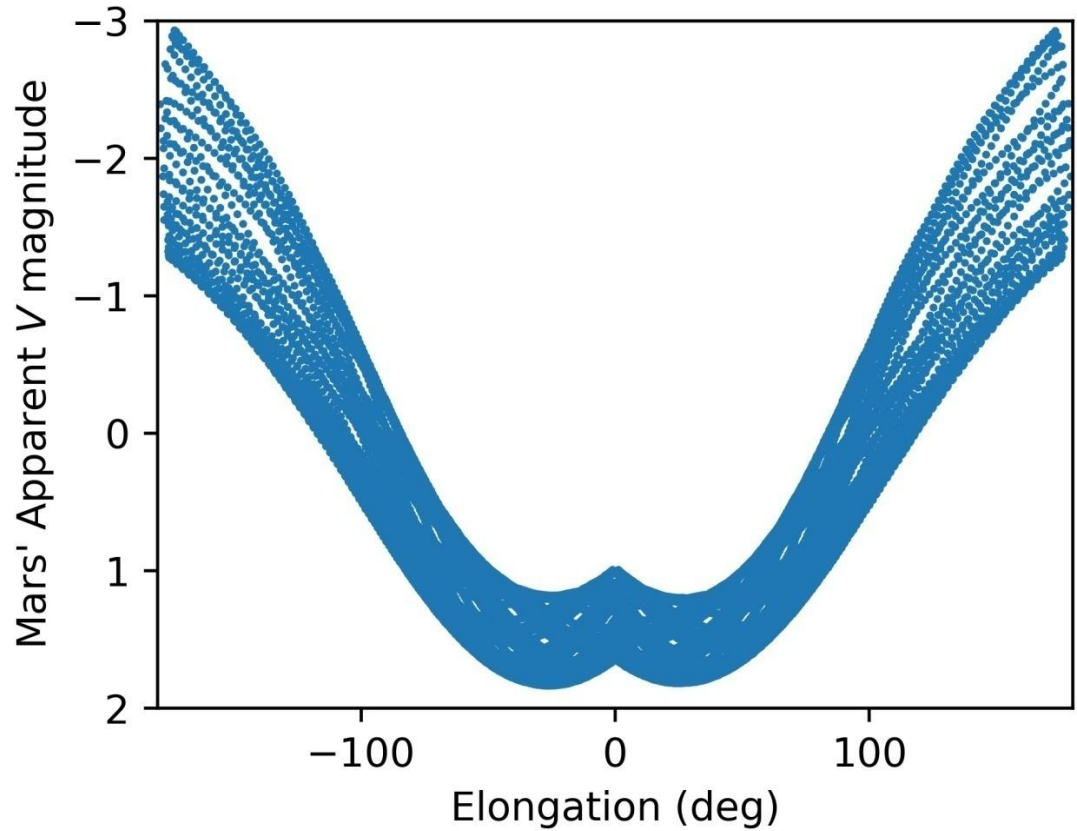

*Fig. 7a. The predicted apparent magnitude of Mars not including the change in magnitude arising from longitude or seasonal changes in surface markings. The 'opposition surge' is wide with a decrease in magnitude beginning about 25$^o$ before conjunction.*

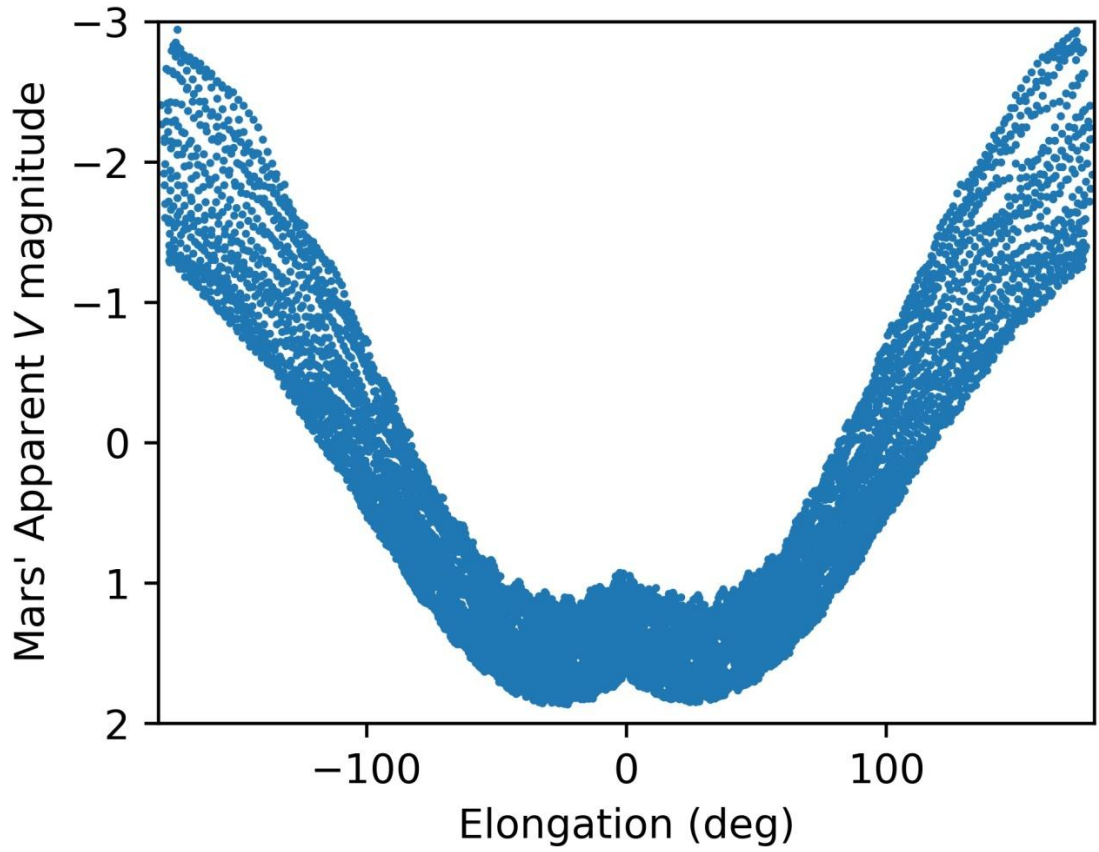

*Fig. 7 b. The contributions of longitude and seasonal effects are included making the curves noticeably noisier. But the surge is still noticeable near conjunction.*

4.5 Jupiter's magnitude statistics

The orbital period of Jupiter is 11.86 Earth years, so 5 jovian years were required to generate at least 50 years of magnitudes. In order to sample an integer number of synodic periods, the selected time span slightly exceeded 60 years, beginning on 1986-Dec-20 and ending on 2047-Jan-12. Fig. 8 shows the apparent magnitude with apparent elongation over this time period.

The brightest magnitude computed from Eq. 8, $V$ = -2.94, is predicted to occur on 2034-Oct-01 when Jupiter will be relatively near the perihelion of its orbit and very close to opposition from the Sun. Thus, it is the same magnitude as Mars at its brightest. Furthermore, Jupiter at its brightest is about equal to Venus at its faintest. The faintest magnitude, $V$ = -1.66, occurred on 2016-Sep-26 fairly close to aphelion and with Jupiter on the far side of the Sun. The planet's mean magnitude is -2.20 and the standard deviation is 0.33.

The nominal mean opposition magnitude for Jupiter computed from its semi-major axis in the same manner as that for Mars is -2.70. The mean derived from analysis of the daily magnitude values is also −2.70 and the standard deviation is 0.17. These and other statistics are summarized in the first part of Table 4.

The variation of the apparent magnitudes for Jupiter due to distance exceeds that due to the phase function because of the limited range of phase angles visible from Earth. The second part of the Table indicates 1.27 magnitudes of variation for distance while that for phase angle is just 0.08.

Table 4. Statistics of the $V$ magnitude for Jupiter

```
          Apparent magnitudes

Brightest                              -2.94
Faintest                               -1.66
Mean                                   -2.20
Standard Deviation                      0.33
Standard Deviation of the Mean          0.00
Mean Opposition                        -2.70

          Variations by component

Earth-Jupiter distance                  1.06
Sun-Jupiter distance                    0.21
Both distances                          1.27
Phase angle                             0.08
```

```
Magnitudes evaluated            21,937
Start date                      1986-Dec-20
Stop date                       2047-Jan-12
```

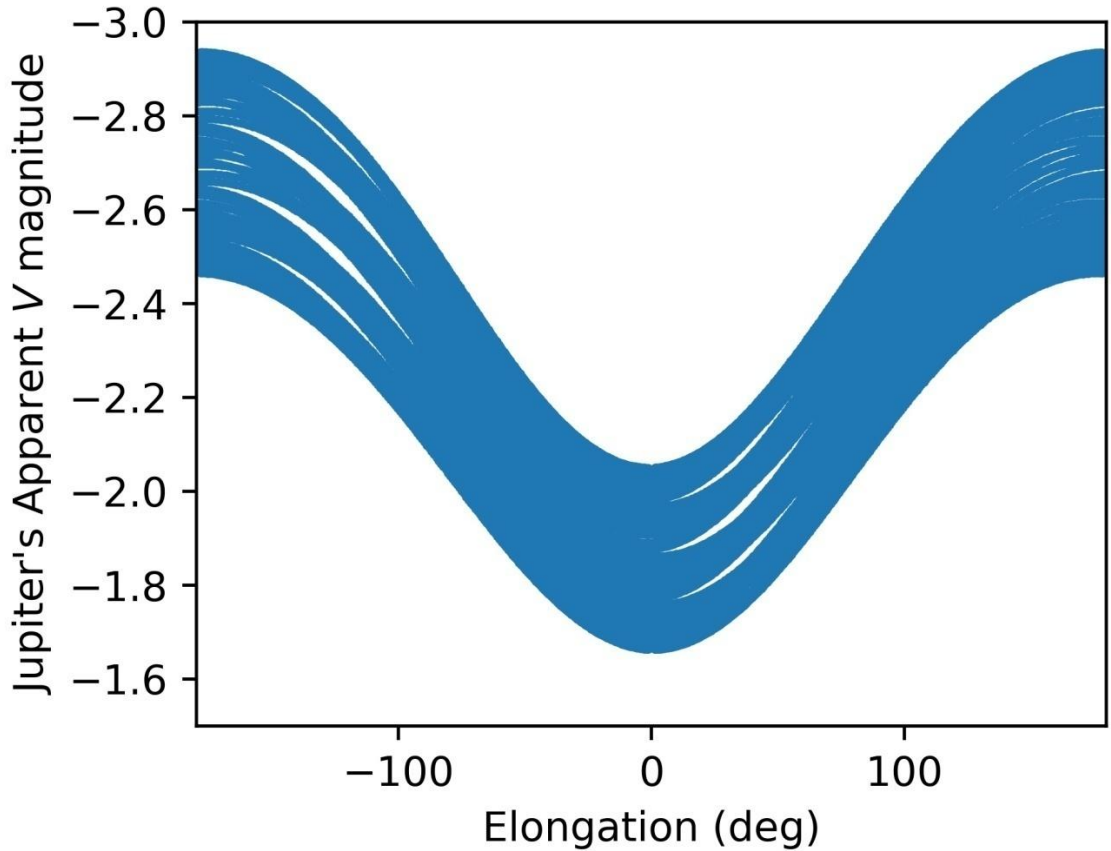

*Fig. 8. The width of the band for the predicted apparent magnitude of Jupiter is mainly caused by the differences in Earth-Jupiter and Jupiter-Sun distance at similar elongations. The structure in the bands is likely an artifact of the near 12:1 commensurability between the Earth's orbital period and Jupiter's synodic period with the Earth.*

4.6 Saturn's magnitude statistics

The orbital period of Saturn is 29.46 years; so, 2 Saturnian years are needed to cover 50 years. In order to sample an integer number of synodic periods as well, a time span of 59 years, beginning on 1987-Jun-30 and ending on 2046-June-30 was selected. All the geometrical aspects including the varying inclination angles of the rings to the Sun, $β_S$, and to the Earth, $β_E$, and their combined value, $β$, as indicated in Eq. 10 were taken into account. Fig. 9 shows Saturn's predicted apparent magnitudes.

The brightest magnitude in the time span examined is -0.55 on 2032-Dec-25 when the rings will be near their maximum inclination, and Saturn is simultaneously near the perihelion of its orbit and at opposition from the Sun. The faintest would be 1.17 on 2025-Apr-20 when the rings are nearly edge-on.

The mean magnitude of Saturn is 0.46. Generally, the standard deviation is less for planets that are further from the Earth because the quantity *log ( r d )* is closer to being constant. However, the standard deviation of Saturn, 0.34, is somewhat greater than that of Jupiter because of the added variability arising from its rings.

Likewise, the nominal mean opposition magnitude for Saturn cannot be computed from its semi-major axis using the same simplistic method as that for Mars and Jupiter because of the variable influence of its ring system. However, the mean derived from analysis of the daily magnitude values was found to be 0.05 and the standard deviation was 0.33. The statistical results mentioned above are summarized in the first part of Table 5.

The Saturn paper computed the brightest magnitude based on three simultaneous conditions: the planet being at perihelion, the planet being at opposition from the Sun, and the rings having their maximum inclination. Likewise, the faint limit was computed for aphelion at conjunction with the Sun while the rings were edge-on. However, those conditions are not met in the current era. In particular, maximum and minimum $β$ angles are intermediate between perihelion and aphelion. Therefore, the actual range of magnitudes is less than that given in the Saturn paper.

Saturn is the first planet in order from the Sun where the variation of its apparent brightness due to distance is less than one magnitude. The variations are limited by the relatively small eccentricity of the planet's orbit (*e* = 0.056) and its great distance from the Earth. The second part of the Table lists the variations due to the Earth distance, the Sun distance and the combined effect.

The variation caused by the phase angle cannot be deduced from the daily magnitudes because $α$ appears along with $β$ in the final term of Eq. 10. Therefore, the equation was solved explicitly for variations due to *α* and *ϐ* separately. That is, the magnitude difference from the minimum and maximum values of *α* (0.0 and 6.3°) was evaluated over the full range of *ϐ* (0.0 to 26.8°), and vice-versa. The Table shows that Saturn's magnitude can vary by 0.34 magnitude due to *α* and by 0.99 due to *ϐ*.

Table 5. Statistics of the *V* magnitude for Saturn

```
                Apparent magnitudes

Brightest                              -0.55
Faintest                               +1.17
Mean                                   +0.46
Standard Deviation                      0.34
Standard Deviation of the Mean          0.00
Mean Opposition                        +0.05

                Variations by component

Earth-Saturn distance                   0.69
Sun-Saturn distance                     0.24
Both distances                          0.93
Phase angle                             0.34
Inclination                             0.99

Magnitudes evaluated                   21,551
Start date                         1987-Jun-30
Stop date                         2046-June-30
```

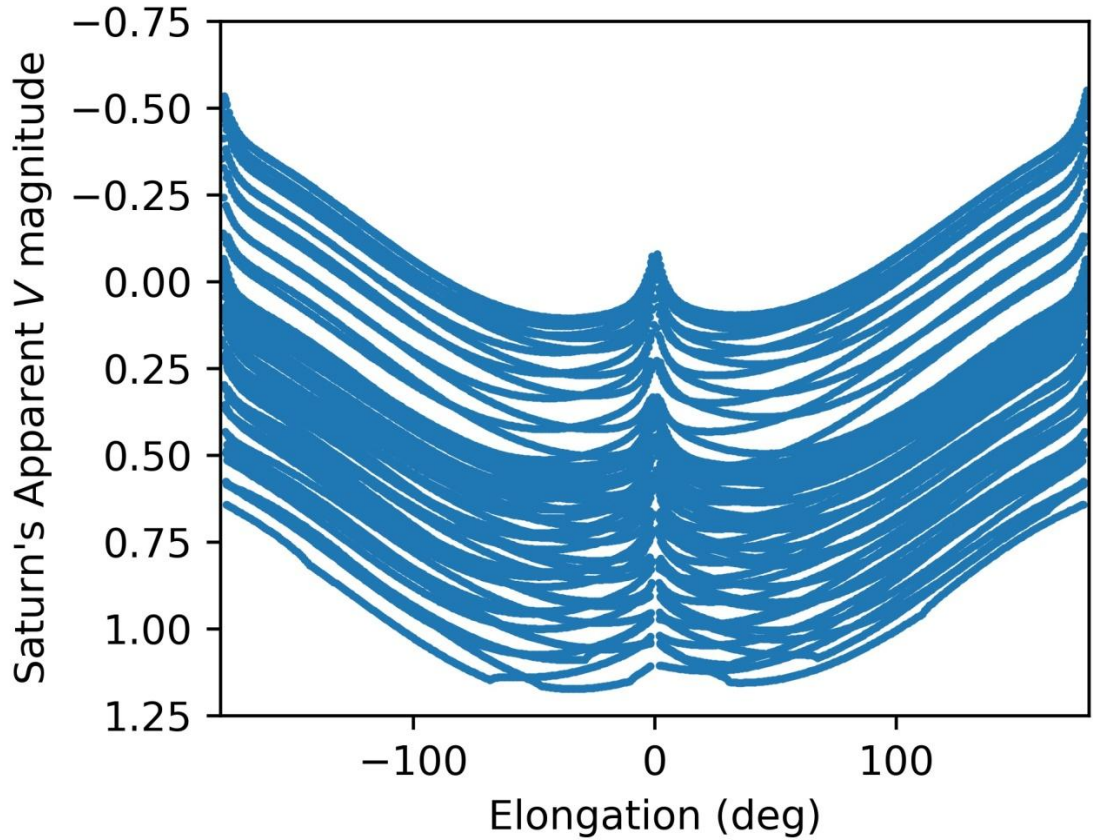

*Fig. 9. The width of the band for the predicted apparent magnitudes for Saturn is caused primarily by the changing tilt of the ring plane with respect to the observer. Saturn has a noticeable brightness surge at both opposition (elongation ~180 $^o$) and conjunction (elongation ~0 $^o$). The size of the surge become more prominent as the inclination of the ring plane with respect to the observer increases in the upper portion of the band.*

4.7 Uranus' magnitude statistics

The orbital period of Uranus is 84.02 Earth years, so less than one Uranian year was needed to cover 50 years. In order to sample an integer number of synodic periods as well, a time span of 85 years beginning of 1974-Jun-27 and ending on 2059-Jul-03 was selected. The geometrical aspects included planetographic sub-Earth latitude, $φ'_E$, the corresponding sub-solar latitude, $φ'_S$ and the combined value, $φ'$, whose effect on V is indicated by Eq. 14.

The brightest magnitude for Uranus will be V = 5.38 on 2054-Mar-29 at opposition. Perihelion occurs about 4 years earlier but at that time $φ'$ will be very small and so its predicted brightness is suppressed. By 2054 $φ'$ increases to about $20^o$, which enhances its brightness. The faintest magnitude, 6.03, occurred on 2008-Mar-09 when the planet was near conjunction with the Sun, was near aphelion and the sub-latitude value was only about one degree.

Fig. 10 shows the predicted change in Uranus' apparent magnitude with elongation from the Sun. The mean apparent magnitude of Uranus is 5.68 with a standard deviation of 0.17. The mean opposition magnitude of Uranus is complicated because of the dependence of V on $φ'$. So, the actual values of V were computed for the 84 oppositions that occurred over the time span of the analysis. The mean was 5.57 and the standard deviation was 0.15. The magnitude statistics for Uranus are summarized in the first part of Table 6.

Uranus is so far removed from the Sun and the Earth that the combined effect of their distances on the planet's apparent brightness is only 0.54 magnitude. The variation due to $φ'$ is 0.07 magnitude and that due to the phase angle is essentially zero. The second part of the Table lists these variations.

Table 6. Statistics of the *V* magnitude for Uranus

```
                  Apparent magnitudes

Brightest                                 5.38
Faintest                                  6.03
Mean                                      5.68
Standard Deviation                        0.17
Standard Deviation of the Mean            0.00
Mean Opposition                           5.57

                 Variations by component

Earth-Uranus distance                     0.43
Sun-Uranus distance                       0.21
Both distances                            0.54
Phase angle                            < 0.01
Sub-latitude                              0.07

Magnitudes evaluated                    31,044
Start date                          1974-Jun-27
Stop date                           2059-Jul-03
```

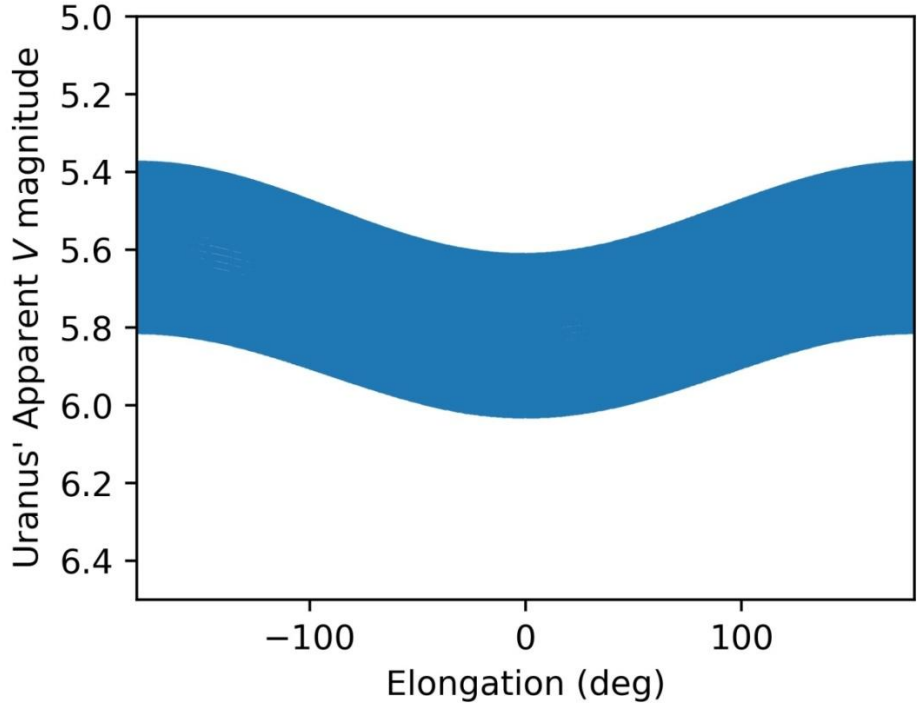

*Fig. 10. The predicted apparent magnitude of Uranus varies with the planet's sub latitude and its distances from the Earth and Sun.*

4.8 Neptune's magnitude statistics

The intrinsic variability of Neptune was characterized in section 3.8. The planet was faintest prior to 1980 and then gradually brightened until 2000 when it reached a plateau. Therefore, the time span selected for the analysis differs from that for the other planets in two ways. One difference is that there are three time segments as explained below. The other is that they are not centered at 2017.0. The choice of time spans allows for statistics to be computed for the planet's intrinsic brightness both at the present time and also when it was fainter.

The first time segment consists of the 3-year period from 1958-Jan-09 until 1961-Mar-28, which includes the last aphelion before the brightening occurred. The second is another 3-year period from 2040-Jan-01 through 2043-Dec-31 which includes the first perihelion after the recent brightening. The third is a period of slightly more than 165 years beginning on 2000-Jan-01 and ending on 2165-Jan-03. This spans the 164.8-year orbital period of Neptune during the plateau with extra days added to equal 164 integer synodic periods. The three time-dependent parts of Eq. 16 were applied in accordance with the dates being evaluated.

The brightest magnitude during brightness plateau, in Fig. 11, will be 7.67 on 2042-Oct-31. Neptune will then be near perihelion and near opposition from the Sun. This prediction depends on the brightness of Neptune remaining constant at its post-2000 level. The faintest magnitude before the brightness increase was 8.00 on 1959-Oct-30, which was near aphelion and near solar conjunction. The magnitude extremes, which are listed in the first part of Table 7, take into account the variability of Neptune as it is currently understood.

The other statistical quantities, which pertain to the brightness of Neptune at present, were determined by analysis of the 165-year period beginning after the brightness plateau commenced. (A two-day time spacing was used instead of daily spacing because of the large number of data points.) The standard deviation of 0.06 magnitude is the smallest for any planet. The mean opposition magnitude, 7.71, was the same whether determined from analysis of the bi-daily magnitudes or computed from the semi-major axis of Neptune's orbit.

The magnitude variation components are listed in the second part of the Table. The combined effect of the Sun-Neptune and Earth-Neptune distances on the apparent brightness of this most distant planet is only 0.22 magnitude. The 'intrinsic' variation refers to the brightening of Neptune between 1980 and 2000.

Table 7. Statistics of the *V* magnitude for Neptune

```
                Apparent magnitudes

Brightest                               7.67
Faintest                                8.00
Mean                                    7.78
Standard Deviation                      0.06
Standard Deviation of the Mean          0.00
Mean Opposition                         7.71

                Variations by component

Earth-Neptune distance                  0.18
Sun-Neptune distance                    0.04
Both distances                          0.22
Phase angle                           < 0.01
Intrinsic                               0.11

Magnitudes evaluated                  30,132
Start date                       1958-Jan-09
Stop date                        2165-Jan-03
```

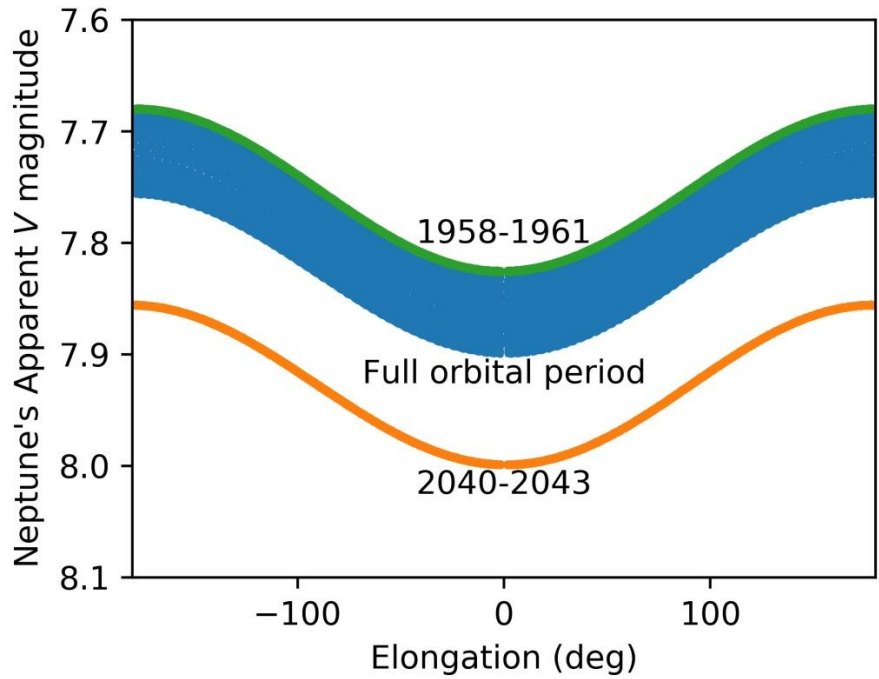

*Fig. 11. The apparent magnitude predicted for Neptune. The band shows the change in its apparent magnitude arising from the Sun-Neptune and Neptune-Earth distances. The curves at the extremes arise from the change in Neptune's intrinsic brightness.*

4.9 Extreme magnitude statistics

This section highlights some of the extremes of the apparent magnitudes of the planets and of their variations. Venus is the brightest planet at all times, even when it is transiting the Sun. Mars and Jupiter attain the brightest opposition magnitudes. Neptune is the faintest planet with the possible exception of Mercury transiting the Sun. The apparent magnitudes of Mercury and of Neptune have the greatest and least standard deviations, respectively. The extremes mentioned above are listed in the first part of Table 8. The ranges of apparent magnitudes are shown in Fig. 12 as a function of elongation and in Fig. 13 in order of the planets' distances from the Sun.

The extremes of the variations due to individual components of the magnitude equations are listed in the second part of the Table. The Earth-planet distance effects Venus most strongly and likewise with the combination of Earth- and Sun-distances. Mercury experiences the greatest variation due to Sun-planet distance because its orbit is the most eccentric of all the planets. Phase angle also affects Mercury most strongly. The remaining five components in the second part of the Table (rotational, orbital, inclination, sub-latitude and intrinsic) are unique to each of the planets listed.

The third part of the Table indicates that the apparent magnitude of Neptune experiences the least variations due to distances and due to phase angle. The sum of the number of magnitudes evaluated for all the planets (except the Earth) is given in the final part of the Table.

The extremes summarized above are for geocentric magnitudes. In closing this discussion a few other extremes of interplanetary brightness may be of interest. When Venus is at opposition from the Sun as seen from Mercury its apparent magnitude can be as bright as $V = -8.0$. That is 18 times the maximum brightness of Venus as seen from the Earth. At the other extreme, while the apparent magnitude of Mercury was computed to be +7.2 for 2029-May-13 as seen from the Earth $\alpha = 179.13^{o}$, it would be nearly 3000 times fainter at $V = +15.9$ from the mean distance to Neptune at the same phase angle.

Table 8. Extremes of the geocentric *V* magnitude for all planets

```
                    Apparent magnitudes

Brightest                         Venus         -4.92
Brightest Mean                    Venus         -4.14
Brightest Opposition              Mars          -2.94
Brightest Opposition              Jupiter       -2.94
Brightest Mean Opposition         Jupiter       -2.70
Faintest                          Neptune        8.00
Faintest (in principle)           Mercury           *
Faintest Mean                     Neptune        7.78
Faintest Mean Opposition          Neptune        7.71
Greatest Standard Deviation       Mercury        1.78
Least Standard Deviation          Neptune        0.06

               Greatest variations by component

Earth-planet distance             Venus          4.08
Sun-planet distance               Mercury        0.91
Both distances                    Venus          4.09
Phase angle                       Mercury       10.82
Orbital                           Mars           0.15
Rotational                        Mars           0.11
Inclination                       Saturn         0.99
Sub-latitude                      Uranus         0.07
Intrinsic                         Neptune        0.11

                 Least variations by component

Earth-planet distance             Neptune        0.18
Sun-planet distance               Neptune        0.04
Both distances                    Neptune        0.22
Phase angle                       Neptune      < 0.01

Magnitudes evaluated                          165,990

* Mercury may be fainter than Neptune when it is
transiting the Sun.
```

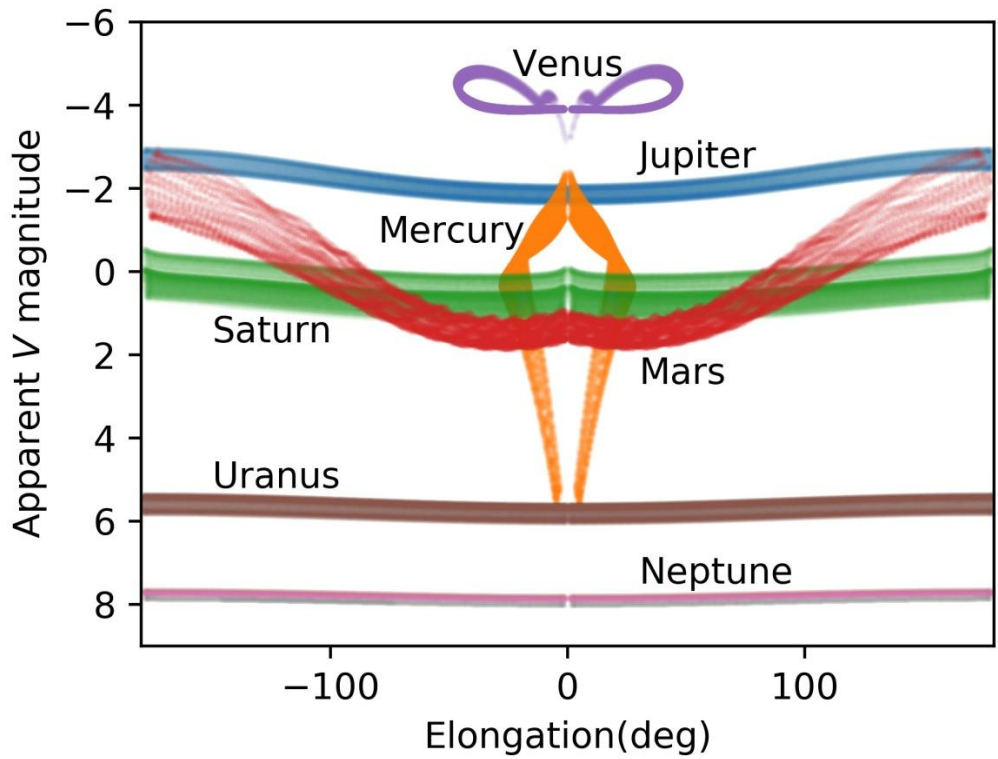

*Fig. 12. The magnitude-versus-elongation graphs from sections 4.1 through 4.8 are combined to illustrate their similarities and differences.*

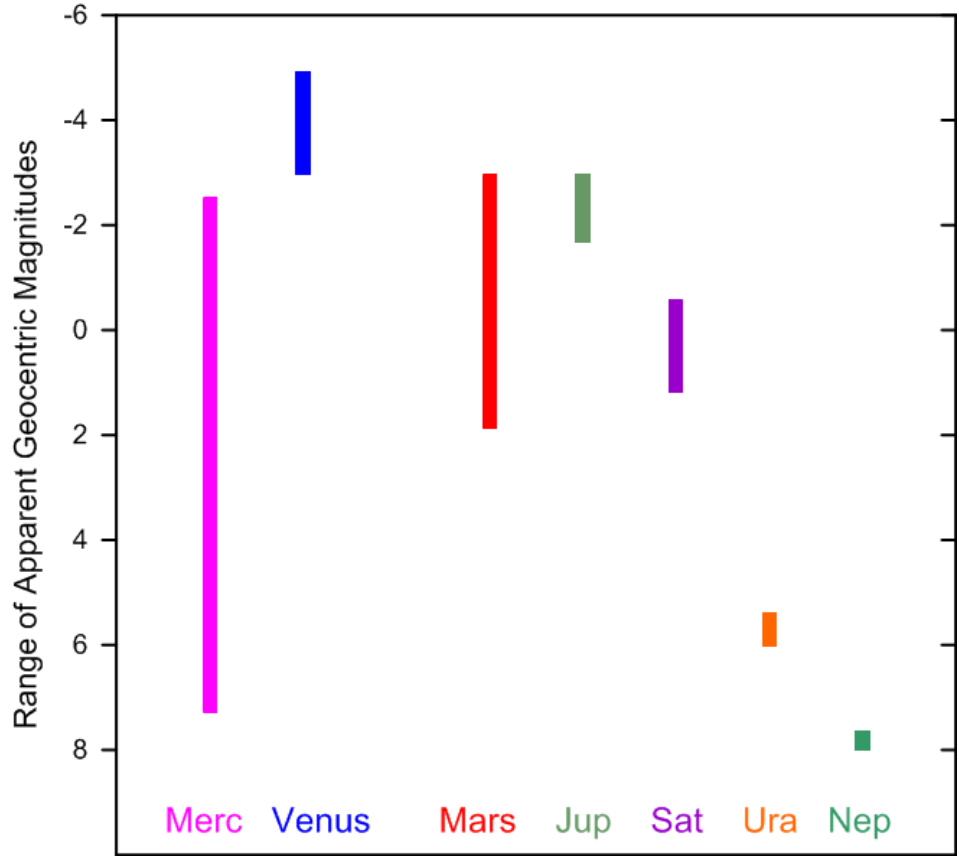

*Fig. 13. The magnitude ranges shown in order of the planets' distances from the Sun. Those of the terrestrial planets tend to be wide and overlapping. Those of the giant planets are generally narrower, do not overlap and increase monotonically with distance from the Sun.*

5. Other band-passes

The previous sections of this paper focused on apparent *V*-band magnitudes. However, there are 10 other wide band-passes commonly in use. Apparent magnitudes and albedos in these bands may be useful to observers studying or recording data at those wavelengths. The total of 11 band-passes are associated with two separate photometric systems. The older one was defined by Johnson and Morgan (1953) and later modified by Cousins (Cousins, 1976a and 1976b). That system spans the spectrum from about 0.3 to 1.0 µm with 7 discrete band-passes including *V*. Table 9 lists their specifications. The Johnson-Cousins system is the traditional wide-band photometric system.

Table 9. Wavelengths and widths of Johnson-Cousins band-passes

| Filter | Effective Wavelength (µm) | Full-width half-max. (µm) |
|---|---|---|
| U | 0.360 | 0.068 |
| B | 0.436 | 0.098 |
| V | 0.549 | 0.086 |
| R | 0.700 | 0.209 |
| I | 0.900 | 0.221 |
| $R_C$ | 0.641 | 0.158 |
| $I_C$ | 0.798 | 0.154 |

Meanwhile, the Sloan photometric system (Smith et al. 2002) is becoming the new standard for astronomical photometry. The five primary bands (u, g, r, i, and z) of the Sloan Digital Sky Survey cover approximately the same total range of wavelengths as do the Johnson-Cousins system but the individual band-passes differ. Smith et al. established a system of 158 standard stars with Sloan magnitudes, which is preferred for photometry. Those magnitudes are indicated with primes (that is, u', g', r', i', and z') and the specifications are given in Table 10.

Table 10. Wavelengths and widths of Sloan band-passes

| Filter | Effective Wavelength (µm) | Full-width half-max. (µm) |
|---|---|---|
| u' | 0.355 | 0.063 |
| g' | 0.469 | 0.143 |
| r' | 0.616 | 0.140 |
| i' | 0.748 | 0.149 |
| z' | 0.893 | 0.117 |

The Sloan band-passes improve upon those of the Johnson-Cousins system in several ways. For one, while the Johnson-Cousins filters are just colored glass, the Sloan filters have a dielectric coating to steepen the shoulders of the response curves. Therefore, the Sloan filters have more rectangular response characteristics than the Johnson-Cousins filters as shown in Fig. 14. A significant advantage of Sloan magnitudes themselves is that they are directly related to absolute flux, that is, they are on the AB system of Oke and Gunn (1983). So, fluxes can readily be derived from magnitudes, and fluxes corresponding with different filters can be compared or combined using simple arithmetic. A final advantage of the Sloan system is that the terrestrial emission line at 0.558 μm lies between the g' and r' filters while it is near the center of the *V* filter on the Johnson-Cousins system. The SDSS, the Panchromatic Survey Telescope and the Rapid Response System (Pan-Starrs) and the Large Synoptic Survey Telescope (LSST) are all on the Sloan photometric system. Other new photometric instruments are generally incorporating Sloan filters, too.

Mallama et al. (2017) reported reference $M_1(0)$ magnitudes along with geometric albedos and phase curves for all 8 planets in all 11 wide band-passes. Most of the reference magnitudes on the Johnson-Cousins system were taken directly from individualized photometric studies for each planet. (These were referred to as 'the Mercury paper', 'the Venus paper' etc. in Section 3.) Wherever photometric information was missing, reference values were assigned from synthetic magnitudes derived in that study from published spectrophotometry. Photometric data were given preference because they were derived from numerous instruments over long periods of time. On the other hand, synthetic magnitudes were often derived from just one instrument at one epoch. Very little Sloan photometry is available for the planets. Therefore, their reference values were taken to be the averages of Sloan photometry, synthetic magnitudes and values transformed from Johnson-Cousins photometry. Geometric albedos were computed for each band and for each planet. The consistency between Johnson-Cousins and Sloan albedos shown in figure 1 of that paper validates the albedo and magnitude results.

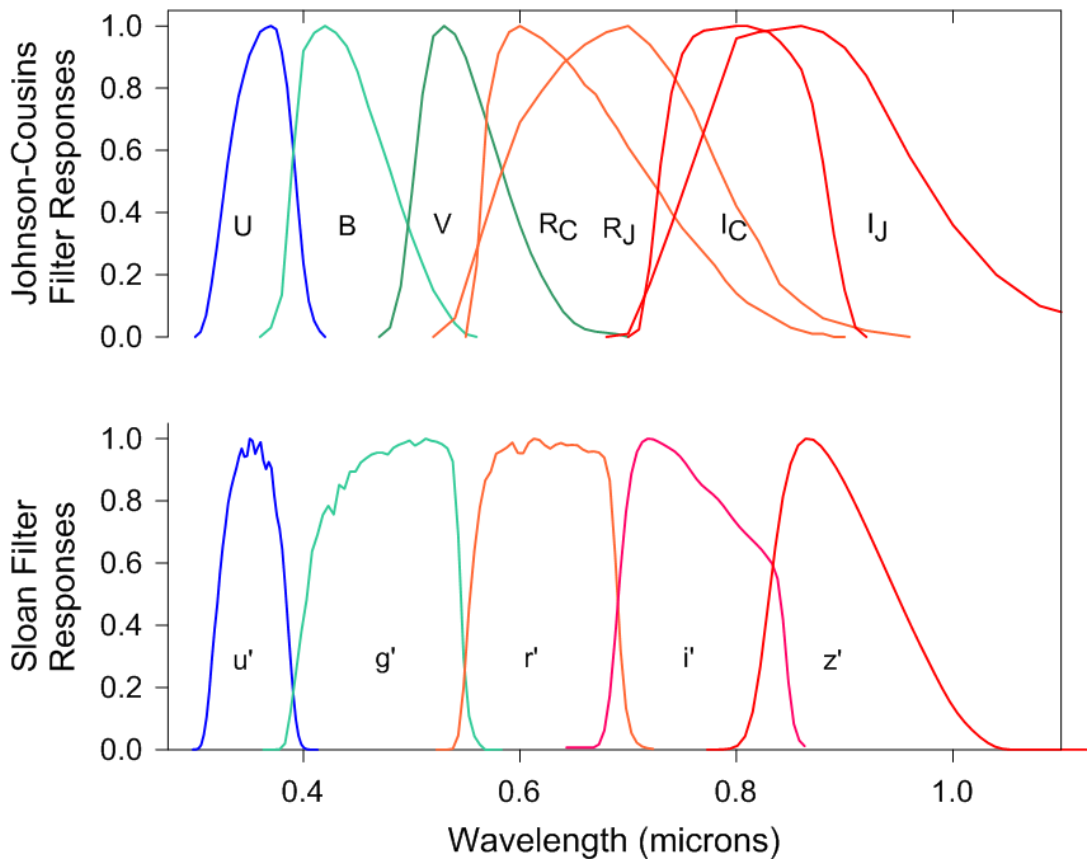

*Fig. 14. Normalized response functions on the Johnson-Cousins and Sloan systems. The Sloan filters generally represent their wavelength intervals more uniformly and overlap their adjacent bands less as compared with those of the Johnson-Cousins system.*

6. U.S. Naval Observatory Products

The apparent *V* magnitudes of all of the planets are tabulated to the nearest 0.1 mag in the U.S. Naval Observatory's products available for general public distribution: *The Astronomical Almanac*, *Astronomical Phenomena*, the Astronomical Applications Department of the U.S. Naval Observatory's web site, and the *Multiyear Interactive Computer Almanac (MICA)*. The apparent magnitudes of Venus, Mars, Jupiter, and Saturn are tabulated in *The Nautical Almanac* and *The Air Almanac*. *The Air Almanac*, however, tabulates only the three planets best placed for navigational observations on any given day. All these products are designed for planning and identification purposes, so only the *V* band with limited precision is included in them.

7. Summary and conclusions

Equations for the apparent visual magnitudes of all eight solar system planets are given. The geocentric magnitudes for each body have been evaluated for a period of at least 50 years in order to characterize the planet's apparent geocentric brightness and its variation.

For Mercury, a sixth-order polynomial is used to model its opposition surge and the sharp magnitude increase at large phase angles. The geocentric brightness of Mercury varies more than any other planet. Two separate equations are needed to model the magnitude of Venus due to a discontinuity at large phase angles. Venus is always the brightest planet as seen from Earth.

The Earth itself does not have a geocentric magnitude. However, its brightness as seen from space has been derived from space-based data and a physical model of its phase curve.

The brightness of Mars depends on its rotational angle and its orbital position in addition to distance and phase angle. Mars and Jupiter are the geocentrically brightest planets after Venus. Terrestrial observations for Mars are limited to phase angles under $50^o$. The magnitude formula for greater angles is an approximation derived by averaging the brightness of a partly-cloudy Earth with that of airless Mercury. The equation for large phase angles is valid to $120^o$.

The magnitude of Jupiter is modeled with a second-order polynomial for the limited range of geocentric phase angles. The formula for larger angles is based upon published observational results from the Cassini spacecraft and is reliable to $130^o$.

Three formulas are given for the apparent magnitude of Saturn. The first two, which cover the range of geocentric phase angles, are for Saturn with its ring system and for the planet's globe alone. The ring system can increase the brightness of Saturn by nearly one magnitude. The third equation is valid for the globe alone up to $\alpha = 150^o$ and is based on published Pioneer spacecraft data.

The geocentric magnitude of Uranus depends on the sub-Earth and sub-solar longitudes with no practical dependence on its very small geocentric phase angle. A supplementary formula for angles up to $154^o$ is based on published Voyager spacecraft results.

The brightness of Neptune has increased over time with a notable rise between 1980 and 2000. Therefore, the equation of its apparent magnitude is time-dependent. Like Uranus, geocentric phase angles have practically no effect on the planet's brightness. Also, like Uranus, the supplementary formula for large phase angle is based on Voyager spacecraft data taken from the literature. That formula is valid up to $133^{o}$.

In addition to the *V* band-pass there are 10 other wide-band photometric intervals covering the spectrum from 0.3 to 1.0 μm. The availability of equations for apparent magnitude in those intervals was briefly discussed.

Finally, planetary magnitude data products available from the U.S. Naval Observatory besides those in *The Astronomical Almanac* are listed. These include *Astronomical Phenomena*, the *Multiyear Interactive Computer Almanac (MICA)*, *The Nautical Almanac* and *The Air Almanac*.

Appendix

Source code for computing planetary magnitudes according to the equations and conditions given in this paper is available on-line. The purpose of the code is to provide verified subroutines applicable to each planet. These may be modified by other scientists and programmers to suit their own applications. The code is described in this appendix. The input and output data files for testing the compiled application program or any derived application are also described. These files are available as supplementary material for this paper. The project, including code and data files, can also be found at https://sourceforge.net/projects/planetary-magnitudes/ under the 'Files' tab in the folder 'Ap_Mag_Current_Version'.

The Ap_Mag source code was developed within the Intel Fortran IDE. It consists of a main controller program and subroutines that are called for computing magnitudes for each of the planets. The main program opens the input and output data files. Then it sequentially calls each of the planet subroutines which, in turn, read several records of input data for each body, compute the planet's corresponding apparent magnitudes and write the results to the screen and to the output file.

Two kinds of checking are performed during program execution. Each subroutine counts the number of discrepancies between pre-computed magnitudes from the input file and those computed by the application itself. The subroutines also check whether the input phase angles are within the observed limits for that planet as indicated in this paper and outputs warning messages if they are not.

Finally, the main program calls an error summary subroutine which displays the number of magnitude discrepancies counted for each planet and the total over all planets. These are categorized as discrepancies greater than 0.01 magnitude and greater than 0.001. When the totals of both these categories are zero, the implementation of the application program has been verified.

In addition to the planet subroutines mentioned above there are two lower level subroutines. One is the Sterling interpolation routine needed for Mars. The other routine converts between planetographic and planetocentric latitude. This conversion may be necessary for either Saturn or Uranus depending on which of those two latitude types (planetocentrir or planetographic) the user is employing.

The input file 'Ap_Mag_Input.txt' contains positional and physical ephemeris data from HORIZONS in addition to pre-computed magnitudes. The data in the various input records has been chosen to test the equations given in this paper under the conditions specified. The record preceding each data record gives the number of the equation being tested along with a brief description of the circumstances. For example, the Neptune section contains records which indicate that they are for times before, during and after the brightness increase that occurred from 1980 through 2000. The input records for Mercury and Venus use the Earth as the observer's location while those for Earth uses Mercury, Venus and Mars. The outer planets use the Earth to test phase angles that are within the geocentric range and use more distant bodies to test phase angles that are beyond the geocentric range.

The output file 'Ap_Mag_Output.txt' contains a record corresponding to each data record in the input file. Each output record compares the pre-computed magnitude from the input file with that generated by the application. The error totals are given for each planet, in turn, and at the end of the file for the sum over all planets.

It should be noted that the Ap_Mag program was written expressly for this paper by author AM. It is *not* the software that will be used to generate magnitudes for the published version of *The Astronomical Almanac* or other products of the Astronomical Applications Department of the U.S. Naval Observatory.

Acknowledgments

The authors thank the anonymous referee for a very thorough and knowledgeable review. The comments and suggestions helped to improve our paper.